\documentclass[aip,pop,
 amsmath,amssymb,
 reprint,
]{revtex4-1}
\usepackage[svgnames]{xcolor}
\usepackage{graphicx}
\usepackage{amsmath}
\usepackage{amssymb, amsthm, mathtools, bm, esint}
\usepackage{physics, siunitx}
\usepackage{hyperref}
\hypersetup{
     colorlinks   = true,
     citecolor    = blue,
     linkcolor    = blue
}
\usepackage{booktabs}
\usepackage{makecell}
\usepackage{amsfonts}
\usepackage{dutchcal}
\usepackage{bm}
\usepackage{color,soul}
\usepackage{array}
\usepackage{pythonhighlight}
\usepackage{todonotes}

\AtBeginDocument{\RenewCommandCopy\qty\SI}
\ExplSyntaxOn
\msg_redirect_name:nnn { siunitx } { physics-pkg } { none }
\ExplSyntaxOff

\usepackage{calligra}

\DeclareMathAlphabet{\mathcalligra}{T1}{calligra}{m}{n}
\DeclareFontShape{T1}{calligra}{m}{n}{<->s*[2.2]callig15}{}

\newcommand{\etal}{$et$~$al$}

\usepackage{tikz}
\usetikzlibrary{calc}
\newlength\thetawidth
\newlength\thetaheight

\newcolumntype{P}[1]{>{\centering\arraybackslash}p{#1}}

\newcommand{\p}[1]{\left(#1\right)}
\newcommand{\fp}[1]{(#1)}
\newcommand{\br}[1]{\left[#1\right]}
\newcommand{\cu}[1]{\left\{#1\right\}}



\newcommand{\diff}{\ensuremath{\operatorname{d}\!}}
\renewcommand{\dv}[2]{\frac{\mathrm{d} #1}{\mathrm{d} #2}}
\newcommand{\dvOne}[1]{\frac{\mathrm{d}}{\mathrm{d} #1}}

\renewcommand{\pdv}[2]{\frac{\partial #1}{\partial #2}}
\newcommand{\pdvOne}[1]{\frac{\partial}{\partial #1}}
\newcommand{\pdvN}[3]{\frac{\partial^{#1} #2}{\partial #3^{#1}}}

\renewcommand{\grad}{\nabla}

\usepackage{graphicx}
\usepackage{dcolumn}
\usepackage{bm}

\usepackage[utf8]{inputenc}
\usepackage[T1]{fontenc}
\usepackage{mathptmx}
\usepackage{etoolbox}
\usepackage{multirow}
\usepackage{tikz} 
\usetikzlibrary{shapes.geometric, arrows} 
\usetikzlibrary{positioning}
\usepackage{xcolor}
\usepackage{soul}

\newcommand{\Pii}{\Pi \hspace{-1.45mm}\Pi}
\newcommand{\ceiling}[1]{\lceil #1 \rceil}
\renewcommand{\vec}[1]{\boldsymbol{#1}}

\makeatletter
\def\@email#1#2{%
 \endgroup
 \patchcmd{\titleblock@produce}
  {\frontmatter@RRAPformat}
  {\frontmatter@RRAPformat{\produce@RRAP{*#1\href{mailto:#2}{#2}}}\frontmatter@RRAPformat}
  {}{}
}%
\makeatother

\makeatletter
\newcommand*\bigcdot{\mathpalette\bigcdot@{1.3}}
\newcommand*\bigcdot@[2]{\mathbin{\vcenter{\hbox{\scalebox{#2}{$\m@th#1\bullet$}}}}}
\makeatother

\begin{document}

\preprint{AIP/123-QED}

\title{Numerical thermalization in $n$-D particle-in-cell simulations}

\author{R. M. Park}
\affiliation{Department of Nuclear Engineering and Radiological Sciences,
University of Michigan, Ann Arbor, MI 48109, USA}
\email{rmpark@umich.edu}

\author{C. H. Moore}
\affiliation{Sandia National Laboratories, Albuquerque, NM 87185, USA}

\author{S. D. Baalrud}
\affiliation{Department of Nuclear Engineering and Radiological Sciences,
University of Michigan, Ann Arbor, MI 48109, USA}

\date{\today}

\begin{abstract}
The particle-in-cell (PIC) simulation method is often understood to solve the
collisionless Vlasov equation due to the finite shape of its macroparticles. In
reality, it can suffer from artificially high collisionality due to the
underresolution of particle number; i.e., the use of a large macroparticle
weight. The degree to which particle shape effects compensate for a large
macroparticle weight in 1D, 2D, and 3D is presented. The collision time is
calculated from PIC simulations based on the decay rate of the velocity
autocorrelation function and compared directly with the kinetic theory of
Okuda, Birdsall, and Langdon. The theory is found to accurately predict the
simulated collision time with varied grid spacings, plasma conditions, and
simulation dimensionalities. The result is a means to predict the timescale of
self-consistent Coulomb interactions in the PIC simulation and thus
characterize the relevance and implications of numerical thermalization as a
function of grid spacing and macroparticle weight. It is determined that
reaching the physical thermalization time, let alone approximating the
collisionless Vlasov limit, may often be intractable in 3D for macroparticle
sizes that resolve the Debye length.
\end{abstract}


\maketitle

\section{Introduction}
\label{intro}
Particle-in-cell (PIC) is a popular simulation technique because it models
the kinetic behavior of device-scale plasmas in a tractable way.~\cite{birdsall,friedman}
This is possible because the introduction of a spatial grid allows for rapid solution of inter-particle forces and because each computational particle in a PIC simulation usually represents many physical particles.
The interpolation of particle density to a grid gives PIC
particles a finite width which is typically on the order of the Debye length.
Since charge screening limits the Coulomb force in a plasma to roughly within a 
Debye length, it is often assumed that the inter-particle forces leading to
Coulomb collisions are not resolved. It is usually expected on this basis that
PIC provides a solution to the collisionless Vlasov kinetic
equation as long as gradients are resolved by the grid.~\cite{verboncoeur,aydemir} 
Collisions, if deemed necessary to include, are then implemented 
using a Monte Carlo collision routine.~\cite{takizuka,nanbu} 

Here, we show that despite the
weakening of inter-particle forces associated with grid interpolation,
representing physical particles as macroparticles with an artificially scaled mass and charge amplifies these forces, and therefore numerical
collisions, significantly. In higher dimensional simulations, it can be
very challenging to achieve a numerical thermalization time $\tau^{\text{num}}$
that is longer than the physical thermalization time. This presents a major
impediment to achieving the correct transport in large-scale PIC simulations,
particularly in 3D. These numerical collisions contribute to the wrong global
evolution of the velocity distribution and thus affect principal quantities of
interest (e.g., emission, breakdown, dissipation, particle loss, \textit{etc}.)
over timescales $t \gtrsim \tau^{\text{num}}$. A goal of the
present work is to validate a theory for estimating $\tau^{\text{num}}$ so that
PIC users can develop constraints addressing numerical thermalization \textit{a priori}.

Knowledge of numerical thermalization in PIC simulations is not
new.~\cite{birdsall,hockney,birdsall2002,turner} Hockney provided an empirical scaling for
numerical collisions in 2D,~\cite{hockneyJOCP1971} and this was recently
elaborated on in 2D~\cite{jubinPOP2024} and extended to
3D~\cite{jubinTHESIS2025} by Jubin~\etal. Early work by Okuda,
Birdsall~\cite{okudaPOF1970} and Langdon~\cite{langdonPOF1970} developed a
kinetic theory of the PIC method, including a collision operator for shaped
macroparticles interpolated to a grid. Here, the kinetic theory is shown to
accurately predict the self-consistent Coulomb collision time in 1D, 2D and
3D PIC simulations. The primary influences on the collision time are the
macroparticle shape, macroparticle weight, and the physical plasma parameter, $N_\mathrm{D}$.
How these influence the overall collision time is shown to be directly
related to an effective inter-particle force for PIC Coulomb interactions. The
finite shape weakens and smooths the force, whereas the macroparticle weight
amplifies it.
By applying the kinetic theory to predict the numerical
collision time, a range of common examples from previous literature are
reviewed in the context of numerical thermalization.

\begin{figure*}
  \centering
  \includegraphics[width=\linewidth]{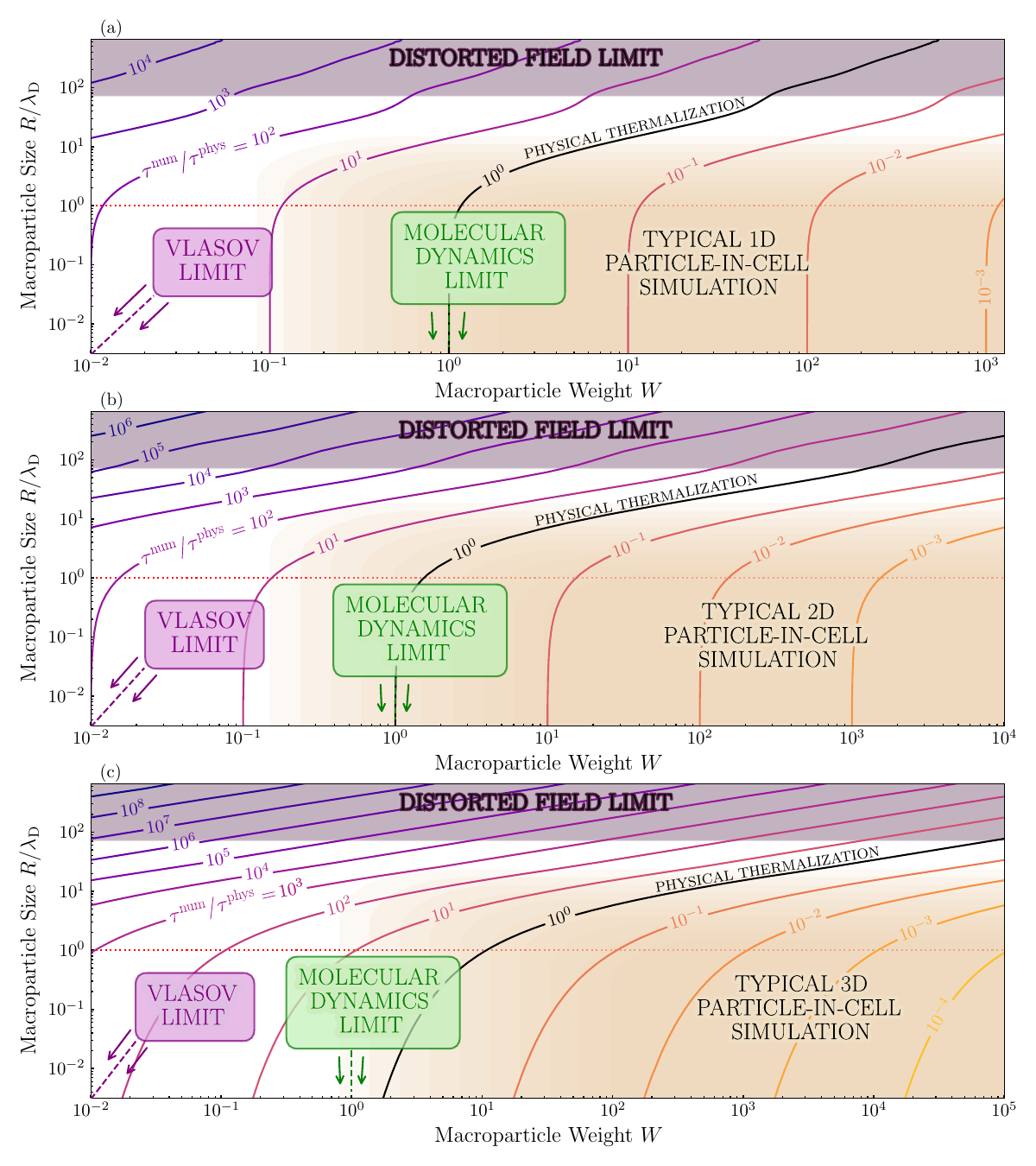}
  \caption{Characteristic regimes of the macroparticle weight and size in (a) 1D, (b) 2D, and (c) 3D. The
  contours obtained from kinetic theory (Eq.~\eqref{big_nu}) show
  numerical thermalization time $\tau^{\rm num}$ normalized by the physical
  thermalization time $\tau^{\rm phys}$. Here, the macroparticle size is half
  the grid width $\Delta$, corresponding to a top hat shape function of width
  $\Delta = 2R$. Simulations with large macroparticle weights suffer from rapid
  numerical thermalization unless they smooth particle density gradients
  and forces over many Debye lengths. In this large $R$ limit,
  numerical thermalization is reduced at the cost of arbitrarily weakening the
  mean (Vlasov) field at length scales shorter than the particle size. Eventually,
  further deformations to the particle shape would violate physical constraints
  on the spatial resolution of the grid, indicated by the purple ``distorted field''
  region. Momentum-conserving PIC algorithms must prevent grid heating and cannot
  operate above $R/\lambda_{\rm D} \gtrsim 1$ (dotted line) unless a filter is used.}
  \label{fig:regimes}
\end{figure*}

The key numerical parameters influencing numerical thermalization are the shape
function $S\p{\vec{r}}$, and the macroparticle weight 
\begin{equation}
      W \equiv N_\text{D}/N_\text{D}^\text{M}.
      \label{eq:weight}
\end{equation}
Here, $N_{\text{D}} \equiv n \lambda_{\text{D}}^3$ is the number of particles
per Debye cube in the physical system being modeled, while
$N_{\text{D}}^{\text{M}} \equiv n^{\text{M}} \lambda_\text{D}^d$ is the number
of macroparticles per Debye element in the $d$-dimensional simulation domain.
In 3D $N_D^\textrm{M}$ is the number of macroparticles per Debye cube, in 2D it
is the number of macroparticles per Debye square, and in 1D the number of
macroparticles per Debye length. The Debye length,
$\lambda_\mathrm{\text{D}}=\sqrt{\epsilon_{0} k_\mathrm{B}T / n q^{2}}$,
is the same in the physical and computational domains regardless of $W$ or $d$.
Given the same computational resources and a domain with characteristic length
$L > \lambda_\mathrm{D}$, a much higher $N_\mathrm{D}^\textrm{M}$
(and thus much lower $W$) is accessible in the lower dimensions. 

In many codes, the weight is defined implicitly, where the user sets a physical density
$n\p{\vec{r}} \sim \br{\text{length}}^{-3}$ and number of macroparticles per grid cell
$N_\mathrm{C}^{\text{M}}$. The weight is then determined from the above definitions by
\begin{equation}
    W = \frac{n\lambda_\mathrm{D}^{3}}{N_\mathrm{C}^{\text{M}}} \p{\frac{\Delta}{\lambda_\mathrm{D}}}^{d},
\end{equation}
where $\Delta$ is the grid spacing. The shape function describes how each macroparticle
will deposit its density onto the grid. During each field solve,
the number assigned to the charge density at grid point $\vec{g} \in \mathbb{Z}^d$ is
\begin{equation}\begin{aligned}
    \rho_{\vec{g}} &= q^{\text{M}}\int_{}^{} \diff^{d} r\ S\p{\vec{r}_{\vec{g}} - \vec{r}} \sum_{j}^{N} \delta^{\p{d}}\p{\vec{r} - \vec{r}_{j}} \\
    &= q^{\text{M}} \left\{S * n_{\delta}^{\text{M}}\right\} (\vec{r}_{\vec{g}}),
\end{aligned}\end{equation}
where $n_{\delta}^{\text{M}}\p{\vec{r}} = \sum_{j}^{N} \delta^{\p{d}}\p{\vec{r} - \vec{r}_{j}}$
is the microscopic distribution of macroparticles and $q^{\text{M}}$ is their charge.
This definition follows that of Birdsall \& Langdon~\cite{birdsall} and Hockney's ``assignment function shape''.~\cite{hockney144}

A summary of the weight and shape parameter space is presented in
Fig.~\ref{fig:regimes}. Contours for the ratio of numerical
and physical collision times computed from kinetic
theory~\cite{okudaPOF1970,langdonPOF1970} show the permissible operating regime
for a simple top hat shape function. The presence of a numerical collision time
which is different than the physical collision time can be understood as the
failure of the macroparticle to represent its physical constituents. Physical
collision times can only be achieved in the region left of the black
curves in the figure. In this region, numerical collisions can be made
negligible, and the physical collision rate can be achieved by introducing a
Monte Carlo collision routine. The effect of $W$ in the limit of small
particle width (and grid spacing), $R\rightarrow 0$, can be understood simply:
the Coulomb collision time in a physical plasma is
$\tau \omega_p \approx N_\textrm{D}/\ln N_\textrm{D}$. In the PIC system,
$N_\textrm{D}$ is replaced by $N_\textrm{D}^\textrm{M}$, decreasing the
collision time by approximately a factor of $W$ relative to the physical value.
Conversely, wider shape functions (or grid spacings) increase the collision
time. 

The quality of the trade-off between shape and weight relies on several assumptions:

(1) The weight must be large enough given the physical $N_\mathrm{D}$ that the
simulation is feasible to run. 

(2) The smallest scale phenomena of interest must be resolved by the shape width.
As macroparticle shapes are made wider, the collisions between
them become weaker. Although this improves numerical thermalization, it has the
well-known side effect of changing the Vlasov mean field and the associated
collisionless processes such as wave dispersion and
damping.~\cite{langdonPOF1970,birdsall2002,hockney,okudaJOCP1972} Therefore any
given application will have a limit (some horizontal line on Fig.~\ref{fig:regimes})
above which the shape size can no longer be reasonably increased. This cutoff
may vary, indicated by the transition to a prohibited ``distorted field limit'' indicated by the purple region.

(3) The resulting numerical thermalization time must be acceptable compared to
the physical thermalization time, the simulation (or residence) time, or
other characteristic timescales, such as electron-neutral collisions.

It may often be the case that (3) is violated in pursuit of (1), or (2)
and (3) are violated unknowingly in higher dimensional simulations where
macroparticles per cell criteria have been established based on convergence
alone.

The kinetic framework for studying properties of PIC simulations emerged from
the work of Okuda, Birdsall~\cite{okudaPOF1970} and
Langdon~\cite{langdonPOF1970} who incorporated the PIC shape function into a
kinetic equation for finite-sized particles. Langdon then considered finite
grid effects in the theory and predicted grid heating and
instability rates~\cite{langdonJOCP1970} by including alias terms that arise
from the field discretization. The particle shape and grid effects were then
verified in 1D simulations by Okuda.~\cite{okudaPOF1972,okudaJOCP1972} These
developments made the basic electrostatic simulation constraints clear: resolve
the Debye length to avoid grid heating and additionally resolve the plasma
frequency to avoid artificial instability. 

Although practical considerations
implemented from the theory are often restricted to instability and
heating phenomena, the question of collisionality and thermalization was also
well known to these authors. Langdon established a scaling for the
characteristic collision times of finite particles in 3D based on their
radius,~\cite{langdonPOF1970} and Okuda listed collisionality reduction as the
motivating reason for studying finite particle effects in the first
place.~\cite{okudaPOF1970} Hockney's empirical model in 2D
PIC~\cite{hockneyJOCP1971} was then developed to emphasize the relationship
between the numerical collision time and the grid dimensions,
\begin{equation}
    \tau^{\text{num}} \omega_{p} \propto N_\mathrm{D}^{\text{M}} + N_\mathrm{C}^{\text{M}},
\end{equation}
where $N_\mathrm{D}^{\text{M}} \equiv n^{\text{M}}\lambda_\mathrm{D}^{2}$
is the number of macroparticles per Debye square.  
Here, $\omega_{p} = \sqrt{n q^{2} / \epsilon_{0} m}$ is the plasma
frequency, which is the same in both computational and physical domains.
Evidently, if there are many more macroparticles per cell compared to
$N_\mathrm{D}^{\text{M}}$, the collision time will be greatly extended
(nominally justifying a cheaper, larger macroparticle weight). 
This paper shows that the Hockney empirical relationship is directly connected
to the weight and shape relationship given by the kinetic theory.

In the modern approach to PIC simulation these considerations are manifest
in the requirement that results converge with respect to the number of
macroparticles per cell. Stated alone,
this condition presumes that a sufficiently large
$N_\mathrm{C}^{\text{M}}$ can be reached for a given
grid spacing. When numerical
thermalization is severe, the user may observe a false convergence
with respect to macroparticles per cell, where a few doublings of the
particle count quickly resolve features of the velocity distribution or
noise in the principal result, but really a few hundred times more
macroparticles are required to avoid numerical thermalization.
The order in which convergence is reached (with respect to particles, or the
grid) is also not standard. Incidentally, there is a contour through
which both grid size and particles per cell can be varied to maintain the same,
possibly unphysical, collision time.
For example, to save resources, one may converge many simulations with respect
to particles per cell, then independently converge with respect to the grid
spacing to improve grid effects,~\cite{radtke} perhaps increasing numerical
thermalization.

Another misleading intuition is
the conflation between numerical thermalization and other forms~\cite{aydemir,hu} of
numerical noise. Noise incurred from sampling initial conditions is
distinct from numerical thermalization which is described by
structured correlations between particle and field fluctuations and therefore
the subsequent dissipation rates of non-equilibrium perturbations.
$\delta f$ methods~\cite{kotschenreuther,hu} remove the numerical noise
associated with a predetermined ``background'' distribution $f_{0}$ by solving
its contribution analytically, where $f = f_{0} + \delta f$. Numerical
thermalization is dependent on the $\delta f$ term itself. It is directly
related to the field fluctuation
$\ev{\delta f \delta\vec{E}}$
produced by $\delta f$ and persists in the
homogeneous $f_{0}=0$ case.~\cite{langdonPOF1979}
Thus, $\delta f$ approaches do address problems with numerical noise, but not
necessarily numerical thermalization.

The paper is organized as follows.
In section \ref{simulation}, the collision
time is validated between electrostatic PIC simulation and kinetic theory in
order to describe how particles exchange momentum and thermalize when the
characteristic width of the macroparticle shape is varied. Section \ref{theory}
introduces the kinetic theory as it relates to the PIC force. Successive
approximations are applied to the modified Lenard-Balescu kinetic equation
to derive simple forms for the thermalization time. Finally, in section
\ref{application}, the considerations of the
paper are consolidated into constraints for PIC users to consider when
designing simulations, and some example calculations of the electron-electron
numerical thermalization time are made for a few interesting examples.

\section{PIC simulations}
\label{simulation}
In the Vlasov solution of a uniform distribution of plasma, the self-consistent
field is zero everywhere. The characteristic particle trajectories are
therefore linear in the absence of an external field, and any kind of
thermalization is absent from the evolution.
In PIC simulation of a uniform distribution of plasma, thermalization is
easily visible.~\cite{jubinPOP2024} The fields fluctuate on small length
and time scales, causing particles to deflect and slow down. These processes
are analogous to physical Coulomb collisions, but they are altered
by the macroparticle weight and shape effects. The uniform system is therefore
a good testbed for numerical thermalization. It is directly applicable to a
local region of space in any PIC simulation, as the grid must resolve gradients
well. 

To obtain the numerical collision time, a series of single species electrostatic PIC
simulations were run with varying grid size $\Delta/\lambda_\mathrm{D}$, and
$N_{\rm D}^{\rm M}$.
The step-by-step velocity data of each individual particle
in the simulation was extracted directly from equilibrium
simulation data to calculate the velocity autocorrelation function,
$Z(t)$ (Eq.~\eqref{autocorrelation}).~\cite{hansen} 
As particles deflect and exchange momentum with other particles in the simulation,
they become decorrelated with their own velocities at previous times and $Z(t) \to 0$.
The exponential decay time of $Z(t)$ is equivalent to the overall numerical
Coulomb collision time.~\cite{hansen}

The simulations presented here
include a single charged species with uniform weighting and
a uniform static neutralizing background. This single species case can be
described by the equations of section \ref{theory} with
omission of species subscripts, i.e. 
$n_{s} \to n, \omega_{ps} \to \omega_{p}, \tau_{ss'} \to \tau$, \textit{etc}.
In usual applications of PIC, the physical $N_\mathrm{D}$ is
specified by an initial condition and either the weight $W$
or the number of macroparticles per cell $N_\mathrm{C}^{\text{M}}$ are chosen as
convergence parameters. The variation in $N_\mathrm{D}^\text{M}$ here
is analogous to the variation in $1/W$ for such an application.
For example, a physical system with $N_\mathrm{D} = 3 \cdot 10^{5}$  (or $\ln \Lambda \approx$10)
could be modeled by a simulation with $N_\mathrm{D}^{\text{M}} = 300$
and thus $W=1000$ or likewise by a simulation with
$N_\mathrm{D}^{\text{M}} = 30000$ and $W=10$, \textit{etc}.

Additional simulations include a grid filter that allows
for a wider macroparticle width than the grid spacing and permits the study of
shape effects without encountering grid heating.

\begin{figure}
  \centering
  \includegraphics[width=\linewidth]{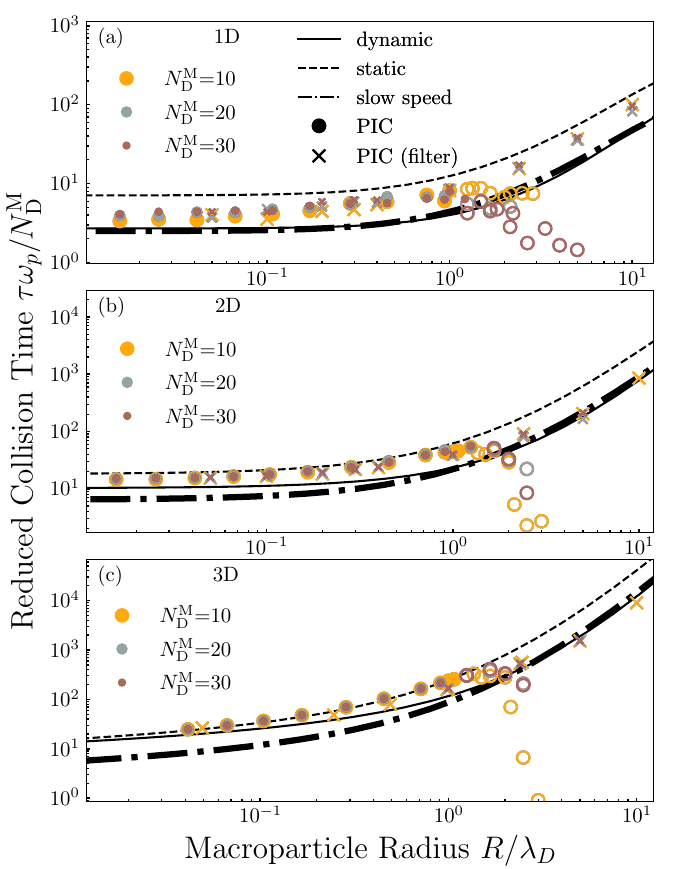}
  \caption{The reduced collision time as computed through the velocity
  autocorrelation of PIC particle data (markers) and the first velocity moment of the
  kinetic Eq.~\eqref{fokker_planck} (lines). The macroparticle
  radius is half the grid spacing $R \equiv \Delta/2$ in the PIC case ($\bigcdot$) and
  half the filter width $R \equiv n_\mathrm{f}\Delta/2$
  in the filtered PIC case ($\times $). The filtered simulations all have $\Delta/\lambda_\mathrm{D} = 0.1$
  and exhibit no grid heating. Unfiltered simulations which exhibited grid heating above 10\% are shown with hollow
  dots ({\large$\circ$}) and do not represent valid measurements of the collision time.
  In 3D, $\tau \omega_P/N_\textrm{D}^\textrm{M}$ is logarithmically dependent on the plasma parameter $\Lambda = 4 \pi N_{\rm D}^{\rm M}$
  in the limit $R \ll \lambda_{\mathrm{D}}$.
  The lines were evaluated using
  Eqs.,~\eqref{nu_slow},~\eqref{nu_static},~\eqref{big_nu},
  and~\eqref{eq:shape}
  where the largest value of $N_{\rm D}^{\rm M} = 30$ was used such that
  $\tilde{k}_\text{max} = 4 \pi N_\mathrm{D}^{\text{M}} = 120 \pi$.}
  \label{fig:tau}
\end{figure}

\subsection{Method}
\label{method}
Particle positions and velocities were stored in
$d$-dimensional vectors and a uniform $d$-dimensional Cartesian mesh was
constructed to represent field values. Periodic boundaries were applied to the
particles and fields. The simulation used a unit system with base length
$\lambda_{\text{D}}$, base time $\omega_p^{-1}$, and base energy
$k_{\text{B}}T$. The thermal speed defined by 
$\bar{v} \equiv \sqrt{k_{\text{B}} T/m} = \omega_p \lambda_{\text{D}}$ defines the
characteristic particle speed, and a target domain length $L/\lambda_{\text{D}}$
was specified alongside $N_{\text D}^{\text{M}}$ for each simulation. The actual
size was constrained such that $N_{\text D}^{\text{M}}$ was exact given the
integer number of macroparticles $N$ rounded up from the target length. The
simulation domain had the same length in each of the $d$ dimensions.

The spatial loading was uniform throughout the domain and velocities
for each degree of freedom were sampled from a normal distribution
with zero mean and unit variance.
Particles were first initialized in a uniform lattice. A lattice
spacing was chosen such that a random number of remainder particles could be
removed to reach the desired total. That is, a periodic lattice with spacing
$l_0 = L/\ceiling{N^{1/d}}$ was generated in the domain.
Then, a random selection of $\ceiling{N^{1/d}}^d - N$ particles were deleted
from the initial configuration to recover the exact $N_\text{D}^{\text{M}}$
specification. It was especially important to use evenly
spaced particle loading in 1D since loading a globally random initial
configuration will overexcite long wavelength modes which are
uncharacteristic of equilibrium and do not decay over the course of the
simulation.~\cite{dawsonRMP1983} Alongside the uniform loading,
a bit-reversed velocity loading recommended by Birdsall~\cite{birdsall}
was sufficient to remove long time correlations that persist in 1D.
Moderate non-equilibrium excitations were also observed in 2D simulations
when a random spatial loading was used. These disappeared with the
uniform loading approach.

After initial particle loading, a second-order explicit leapfrog algorithm
was used to integrate the particle equations of motion by alternating between
particle acceleration and field solver subroutines on each time
step.~\cite{verlet} To interpolate charge density to the grid, many codes
implement a hierarchy of shape functions called the B-splines.~\cite{hockney}
They are generated by the top hat function with the same width as the
grid spacing ($\Delta$) and unit norm
\begin{equation}
    \Pi_{\Delta}\p{r_{\sigma}} =
    \begin{cases}
        1/\Delta & |r_{\sigma}| \leq \Delta / 2, \\
        0, & \text{otherwise.} \\
    \end{cases}
    \label{shape_typ}
\end{equation}
The B-spline shape function of order $n$ is constructed from convolutions of
$\Pi_{\Delta}$,
\begin{equation}
    S^{\p{n}}\p{\vec{r}, \Delta} = \prod_{\sigma}^{d} \Pi_{\Delta}^{\p{* n}} \p{r_{\sigma}}.
    \label{bsplines}
\end{equation}
$\Pi^{\p{*n}}_{\Delta}$ is the normalized top hat distribution convolved with
itself $n$ times. For example,
$\Pi_{\Delta}^{\p{*2}} = \Pi_{\Delta} * \Pi_{\Delta} * \Pi_{\Delta}$.
All simulation data presented here used the $0^{\text{th}}$ order top hat shape
$S\p{\vec{r}} = S^{\p{0}}$ corresponding to the nearest grid
point~\cite{hockney} (NGP) interpolation method. That is, each particle
deposits all of its density on the nearest grid point and subsequently claims
the force of the nearest grid point after the field solve; this method is
momentum-conserving. 
The effects of higher-order shape functions are discussed from the standpoint of the underlying kinetic theory in Sec.~\ref{sec:shapes}. 
Forces were calculated with standard FFT methods and
a finite difference form of Poisson's equation corresponding to
Eqs.~\eqref{Ess}, and~\eqref{Dtensor_App}. 

In some calculations, a filter was applied in order to
increase the effective radius of the macroparticles without coarsening the
grid, thus avoiding grid heating. The filter was implemented in
$\vec{k}$-space as
\begin{equation}
    \varphi\p{\vec{k}} = \frac{\prod_{\sigma}^{d} \text{sinc}(k_\sigma n_\mathrm{f}\Delta/2)
    }{\prod_{\sigma}^{d} \text{sinc}(k_\sigma \Delta/2)}
    \label{ngp_filter}
\end{equation}
where $n_\mathrm{f} \Delta / 2$ is the macroparticle width after the filter is
applied. Note that since the density can only be deposited on the grid points, the
shape can only be widened by an integer multiple of the grid spacing
$n_\mathrm{f}=1,2,3$, \textit{etc}. as in Ref.~\onlinecite{okudaPOF1972}, for
example. In all cases with a filter, the grid width was set to
$\Delta = 0.1 \lambda_{\rm D}$ to avoid heating. 

To begin the evolution, a simple thermostat was applied in order to reach
equilibrium at the appropriate temperature set by $\bar{v}$. The thermostat
was implemented as follows: particle and field equations were integrated for
two plasma cycles of period $\tau_{p} \equiv 2 \pi / \omega_{p}$, during which
the total kinetic energy~\cite{birdsall} of particles $j=1 \dots N$,
\begin{equation}
    E_{\rm kin}(t) = \frac{m}{2}\sum_j^N~\vec{v}_{j}(t-\Delta t/2) \cdot \vec{v}_{j}(t + \Delta t/2)
\end{equation}
was averaged. Then all macroparticle velocities were instantaneously scaled by
\begin{equation}
    \vec{v}_j \to \sqrt{\frac{\frac{d}{2} N k_\mathrm{B}T}{E_{\rm kin}}}\vec{v}_j.
\end{equation}
This process was repeated for 20$\tau_{p}$ to allow for the spatial relaxation
between particles. Next, the thermostat was turned off and an equilibrium
simulation was performed for ${t_\mathrm{sim}=\ 200\omega_{p}^{-1}}$. Particle velocity data was
stored periodically, about every 0.5 $\omega_{p}^{-1}$.

The velocity data was used to calculate the velocity autocorrelation
function (VACF) defined as
\begin{equation}
    Z\p{t} = \frac{1}{d}\ev{\vec{v}\p{{0}} \cdot \vec{v}\p{t}},
    \label{autocorrelation}
\end{equation}
where $\ev{\cdot}$ denotes the equilibrium ensemble average. Assuming
the system is ergodic, the VACF can be calculated by the discrete time average
\begin{equation}\begin{aligned}
    Z\fp{t_{n}} &= \frac{1}{Nd\fp{N_{t}-n}} \sum_{j=1}^{N}\sum_{m=0}^{N_{t} - n} \vec{v}_{j}\fp{t_{m}} \cdot \vec{v}_{j}\fp{t_{n} + t_{m}} \\
                &= \frac{1}{Nd\p{N_{t} - n}}\sum_{j=1}^{N}\vec{v}_{j} \p{-t_{n}} * \vec{v}_{j} \p{t_{n}}
\end{aligned}\end{equation}
where $t_{n},t_{m} \in 0, \Delta t, 2 \Delta t, ..., N_{t} \Delta t$ are the available
time steps and $*$ denotes the dot product convolution operation. The VACF
was then calculated with FFT methods via the convolution theorem.~\cite{teukolsky} 

The velocity autocorrelation function is a well studied object in statistical
physics, and its decay rate is related to the collision time in the
Chapman-Enskog expansion of the kinetic
equation.~\cite{ferziger_kaper} This connection between the microscopic
simulation variables and kinetic theory is facilitated through the Green-Kubo
formalism.~\cite{hansen} The collision time $\tau$ (derived in
Eq.~\eqref{big_nu} of section~\ref{theory}) is thus related to
the VACF by
\begin{equation}
    Z\p{t} = Z\p{0} e^{-t/\tau}.
\end{equation}
At very early times $t \lesssim \omega_p^{-1}$, $Z(t)$ undergoes quadratic decay,
and at very late times, it gives way to higher-order processes which are not
captured by the linearized kinetic theory. To obtain $\tau$, $Z(t)$ was
calculated using all stored velocity data and an exponential decay time was fit to
$Z(t)/Z(0)$ for times $t \in \br{\tau_p, 5 \tau_{p}}$. In all
three dimensionalities ($d=1,2,3$) the collision time is the principal metric
for numerical thermalization. It describes the rate at which particles slow
down and exchange momentum with their neighbors via thermalizing collisions.
As will be seen in section~\ref{theory},
the velocity drag and diffusion timescales are of the same order and have
approximately the same dependence on $\Delta$.
See a note on 1D thermalization in Appendix~\ref{app:1D_thermalization}.
Finally, all results were converged with respect
to the length of the domain $L$ and the time step resolution $\Delta t$.
The reduced collision time reached convergence at about
$L = \max\p{5 \lambda_\mathrm{D}, 3 R}$ for 2D and 3D simulations. A constant
value of $L = 100 \lambda_\mathrm{D} $ was sufficient to resolve size effects
for all 1D simulations. A timestep of about $\Delta t = 0.05 \omega_{p}^{-1}$
was sufficient to resolve time integration effects for all simulations.

\subsection{Results\label{sec:results}} 
Figure~\ref{fig:tau} depicts the collision time measured from PIC simulations
in comparison with collision times derived from kinetic theory; see Sec.~\ref{theory}. It is expected that the theory becomes accurate when
$\tau^{\text{num}} \propto N_\mathrm{D}^{\text{M}} \gtrsim 1$ is sufficiently large, and
convergence to the expected $1/N_\mathrm{D}^{\text{M}}$ scaling is
demonstrated in all dimensions and grid sizes. It is immediately apparent in
both theory and simulation that the numerical thermalization time only depends
strongly on the macroparticle radius when $R \gtrsim \lambda_{\rm D}$. At these
widths, the unfiltered grids ($\Delta = 2R$) suffer from increasingly large
grid heating and thus the equilibrium conditions assumed in the VACF analysis are
violated. Those values are marked hollow and do not represent accurate
measurements of the numerical thermalization time, but rather demonstrate the
onset of grid heating in unfiltered momentum-conserving solver schemes that do
not resolve the Debye length. In the filtered case, there is no observed heating
and a good agreement with the theory is reached in
both large and small macroparticle radius limits. The filtered 3D data is all
within 25\% of the dynamic screening theory model. In the reduced dimensions,
the PIC results are all between the static and dynamic theory models. It has
been observed in MD simulations that stopping power agrees better with the
static model at moderate speed $v \approx \bar{v}$.~\cite{bernstein} This
discrepancy might be explained by onset of nonlinear coupling in the wake of
colliding particles causing the dielectric response to take some form between the
static and dynamic models presented here. In the following sections, the
kinetic theory based on Langdon, Okuda, and
Birdsall~\cite{langdonPOF1970,okudaPOF1970} is applied to derive each of the
collision time models displayed in Fig.~\ref{fig:tau}.

\section{Kinetic theory}
\label{theory}

The particle-in-cell method has an intuitive interpretation as it is comprised of
particles with charges and masses that behave collectively in a similar manner to
physical electrons and ions. This similarity also benefits the theoretical understanding
of the simulation evolution. Just as one would employ a kinetic equation to study
the properties of a physical plasma, one can do the same with a PIC plasma with
surprisingly few modifications.~\cite{birdsall}
Although specific implementations of the algorithm can differ and alter the
exact form of the PIC kinetic equation, the essential features (macroparticle
shape and weight) are present in all PIC simulations and only
critically modify the macroparticle pair force. In analogy with the traditional
theory, velocity moments of the kinetic equation are calculated to obtain
collisional transport coefficients.
These transport coefficients, specifically the momentum transfer collision
time, are formally related to the decay rate of the velocity autocorrelation
function calculated in Sec.~\ref{method} via
Green-Kubo relations.

\subsection{The PIC pair interaction force}

Here, it is demonstrated that the weight and shape effects relevant to
numerical thermalization can be encapsulated into the macroparticle pair force.

Given a set of macroparticles of species $s$ positioned in a PIC simulation
at $\vec{r}_{j}$, the microscopic density may be expressed as
$n_{\delta s}\p{\vec{r}} = \lambda_{\text{D}}^{d-3}W_{s}\sum_{j}^{N} \delta^{\p{d}}\p{\vec{r} - \vec{r}_{j}}$.
The resulting force acting on an $s'$
macroparticle at position $\vec{r}$ may be written as
\begin{equation}
    \vec{F}_{ss'}\p{\vec{r}} \equiv W_{s'}\frac{q_{s} q_{s'}}{\epsilon_{0}}
    \mathcal{F}^{-1} \vec{K}\p{\vec{k}} \mathcal{F}\ n_{\delta s}. 
    \label{Kdef}
\end{equation}
The force kernel $\vec{K}\p{\vec{k}}$ is the $\vec{k}$-space representation
of the force operator acting on a distribution of
simulation particles. Here $\mathcal{F},\mathcal{F}^{-1}$ are forward and
reverse Fourier transform operators defined as
\begin{equation}
    \mathcal{F}  = \int_{}^{} \diff^{d} r e^{-i \vec{k} \cdot \vec{r}}, \qquad
    \mathcal{F}^{-1} = \int_{}^{} \frac{\diff^{d} k}{\p{2 \pi}^{d}} e^{i \vec{k} \cdot \vec{r}}. 
\end{equation}
In a real electrostatic plasma, the force kernel takes the simple Coulomb form
\begin{equation}
    \vec{K}_{\text{C}}\p{\vec{k}} = -\frac{i\vec{k}}{k^{2}}.
\end{equation}
In a momentum-conserving PIC plasma, the kernel becomes an operator modulated by the
Fourier transform of the shape function, $S\p{\vec{k}, \Delta}$, a gradient
discretization tensor $\mathcal{D}\p{\vec{k}, \Delta}$, and an optional filter
$\varphi\p{\vec{k}, \Delta}$,
\begin{equation}
    \vec{K}_{\text{MC}}\p{\vec{k}, \Delta} = - \frac{i \vec{k}}{k^{2}} \mathcal{D}\varphi^{2} S\ 
    \Pii * S.
    \label{eq:pic_kernel_alias}
\end{equation}
Here $\Pii\p{\vec{k}, \Delta} = \sum_{\vec{p}}^{} \p{2 \pi}^{d}\delta^{\p{d}}\p{\vec{k} - \vec{k}_{\vec{p}}}$
is the Dirac comb representing the spatial aliases
$\vec{k}_{\vec{p}} = \vec{k} - 2 \pi\vec{p}/\Delta$ for all $\vec{p} \in \mathbb{Z}^{d}$
which arise from the finite grid spacing.
The convolution operation applies to everything right of its placement such that, e.g.,
$\Pii * S n_\delta = \sum_{\vec{p}}^{} S\p{\vec{k}_{\vec{p}}} n_\delta\p{\vec{k}_{\vec{p}}}$.
The $\vec{p} \neq \vec{0}$ terms do not conserve energy and are thus responsible
for the common problem of grid heating. Regardless of the shape function used,
aliasing can be made negligible by shrinking the grid spacing $\Delta$ well below the
Debye scale, resulting in a small signal $n_\delta\fp{\vec{k}_{\vec{p}\neq \vec{0}}} \to 0$.
One may also use an energy-conserving~\cite{lewis}
(EC) or fully-implicit~\cite{chen,chacon} algorithm to avoid grid heating, in which
case the force kernel takes a different form (see Appendix~\ref{app:PIC_forces}). Rather than incur grid heating,
explicit EC algorithms lose momentum conservation due to the alias terms.~\cite{langdonJOCP1973}

It is presumed in the remaining discussion that avoidance of these aliasing
conditions is of general concern and therefore $\vec{p} \neq \vec{0}$ terms are commonly made
negligible by design, either by resolving the Debye length or implementing a filter
$\varphi$ which attenuates the spatial aliases at the expense of
spatial resolution in the fields.
Retaining only the $\vec{p}=\vec{0}$ term results in the kernel vector
\begin{equation}
    \vec{K}_{\vec{p} \neq 0}\p{\vec{k}, \Delta} = - \frac{i \vec{k}}{k^{2}}
    \cdot \mathcal{D} S^{2} \varphi^{2}.
    \label{pic_kernel_finite}
\end{equation}
For a simple finite difference method, the gradient discretization tensor
has Cartesian elements of the form~\cite{birdsall}
\begin{equation}
    \mathcal{D}_{ij} = \delta_{ij}  \frac{\text{sinc}\p{k_{j} \Delta}}{\sum_l\text{sinc}^{2} \p{k_{l} \Delta/2}}.
    \label{Dtensor}
\end{equation}
This tensor represents modifications to the force spectrum due to a first-order finite
differencing scheme applied to the Gauss and Poisson equations.
The approximation $\mathcal{D} \approx 1$ is valid within 8\% when $k \Delta \leq 1$,
which are the dominant modes contributing to the force in explicit schemes
which have eliminated aliasing via the condition $\Delta \lesssim \lambda_\mathrm{D}$.
In making this approximation, the kernel vector simplifies further:
\begin{equation}
    \vec{K}\p{\vec{k}, \Delta} = - \frac{i \vec{k}}{k^{2}} S^{2} \varphi^{2}. 
    \label{pic_kernel}
\end{equation}
This kernel represents the ``cloud'' plasma model of Okuda.~\cite{okudaPOF1972,okudaPOF1970}
It has the interpretation
that the shape function size is constrained by the grid
width, but the dynamics reflect grid-less particles with rigid and continuous charge distributions
given by $W q \varphi\p{\vec{r}, \Delta} * S\p{\vec{r}, \Delta} * \delta\p{\vec{r} - \vec{r}_{j}}$.

For the purposes of comparing with the PIC simulations from Sec.~\ref{sec:results},
a set of simple filters may be defined by
\begin{equation}
    \varphi_{n_\mathrm{f}}\p{\vec{k}, \Delta} = S\p{\vec{k}, n_\mathrm{f} \Delta} / S\p{\vec{k}, \Delta}
\end{equation}
with $n_\mathrm{f} = 1,2,3,...$ . These filters have the action of multiplying
the width of $S$ by some multiple $n_\mathrm{f}$. The force kernel in this
case becomes,
\begin{equation}
    \vec{K}_\mathrm{f}\p{\vec{k}} = - \frac{i \vec{k}}{k^{2}} S^{2}\p{\vec{k}, n_\mathrm{f} \Delta}. 
    \label{pic_kernel_filter}
\end{equation}
This form suggests an analogy between fixing $n_\mathrm{f} = 1$ (i.e. running an unfiltered simulation)
while varying $\Delta$, and fixing $\Delta$ while varying $n_\mathrm{f}$. Either 
situation can therefore be described by the shape function $S\p{\vec{r}, R}$ with macroparticle radius defined as
$R \equiv  n_\mathrm{f} \Delta/2$, as is done in Sec.~\ref{simulation}.
In both filtered and unfiltered cases, the analysis breaks down when
$\Delta \gtrsim \lambda_\mathrm{D}$, but $n_\mathrm{f} \Delta$ can be arbitrarily
large.

This method was employed for the filtered PIC calculations shown in
Fig.~\ref{fig:tau}.
The $\mathcal{D} \approx 1$ approximation is more valid in the filtered case
for the intermediate range of 
$0.1 \lesssim R/\lambda_{\rm D} \lesssim 1$. This is reflected in
Fig.~\ref{fig:tau} where the filtered
data is closer than the unfiltered data to the theory. The deviations in this $R$
regime for the unfiltered data are likely due to finite gradient effects where
$\mathcal{D} \neq 1$.

\subsection{The Lenard-Balescu-like kinetic equation for PIC}

It is useful to begin with some reasonable assumptions:
\begin{enumerate}
    \item The simulation plasma is ideal (weakly-coupled) such
    that $1/N_{\text{D}s}^{\text{M}}$
    is a suitably small expansion parameter. This is a necessary condition
    in collisionless plasmas and is required to avoid artificial correlation
    heating (ACH) in a PIC simulation.~\cite{acciarri} 
    \item Time aliasing effects are small. It is assumed
    that either the plasma frequency is resolved by the time step or an
    appropriate implicit method is used to avoid numerical instability
    and therefore time aliasing.
    \item Space aliasing effects are small. Either an energy-conserving
    method is used or the Debye length is resolved by the grid to avoid spatial
    aliasing and associated grid heating.
    \item The length and time scales of fluctuations are small compared to the
    length and time scales of the analogous Vlasov evolution. Scenarios with shocks
    and severe nonlinear coupling would require separate treatment.
\end{enumerate}

Given these assumptions, the kinetic behavior of a PIC plasma can be modeled by
a Lenard-Balescu-like equation for shaped
macroparticles~\cite{birdsall,touati,balescu}
\begin{equation}
    \label{fokker_planck}
    \dv{f_{s}}{t} = -\pdv{}{\vec{v}} \cdot \br{\vec{A}_{s} f_{s}} + \pdvN{2}{}{\vec{v}} : \br{\mathbb{D}_{s} f_{s}},
\end{equation}
where
\begin{subequations}
\label{eq:A_D_coeff}
\begin{align}
    \label{drag_coeff}
    \vec{A}_{s}\p{\vec{v}} &=
    \sum_{s'}^{}\frac{\omega_{ps '}\bar{v}_{s'}^{2}R_{ss'}^{A}}{N_{\text{D}s'}} \pdvOne{\vec{v}}\cdot \mathbb{I}_{ss'},  \\
    \label{diff_coeff}
    \mathbb{D}_{s}\p{\vec{v}} &=\sum_{s'}^{} \frac{\omega_{ps '}\bar{v}_{s'}R_{ss'}^{D}}{N_{\text{D}s'}} \mathbb{I}_{ss'},
\end{align}\end{subequations}
and $f_{s}\p{\vec{r}, \vec{v}, t}$ is the macroparticle phase space
distribution for species $s$. The left side of Eq.~\eqref{fokker_planck} is a
total derivative including the particle streaming and Vlasov mean field
terms,
\begin{subequations}\begin{align}
    \dv{f_s}{t} &\equiv \pdv{f_s}{t} + \vec{v} \cdot \pdv{f_s}{\vec{r}}
    + \frac{q\bar{\vec{E}}}{m} \cdot \pdv{f_s}{\vec{v}} \\
    \bar{\vec{E}}(\vec{r}) &= \sum_{s'} \frac{q_{s'}}{4 \pi 
    \epsilon_0}\iint \diff^d v \vec{K}\fp{\vec{r}', \Delta} * f_{s'}(\vec{r}', \vec{v}).
    \label{mean_field}
\end{align}\end{subequations}
which is altered by the convolution of $f_{s}$ with the PIC force kernel operator
$\vec{K}\p{\vec{r} ', \Delta} \equiv \mathcal{F}^{-1} \vec{K}\p{\vec{k}, \Delta} \mathcal{F}$.
The right side of Eq.~\eqref{fokker_planck} describes
the velocity drag~\eqref{drag_coeff} and diffusion~\eqref{diff_coeff} processes
which arise from discrete particle interactions (i.e. collisional effects
first-order in $1/N_{\text{D}s}$). These coefficients depend on the symmetric
tensor given by
\begin{equation}
    \mathbb{I}_{ss'}\p{\vec{v}} = 2 \pi W_{s '}\iint_{}^{} \diff^{d} \tilde{v}' \frac{\diff^{d} \tilde{k}}{\p{2 \pi}^{d}}
    \tilde{\vec{K}}_{\text{S}}\vspace{-1mm} \tilde{\vec{K}}_{\text{S}} \delta\fp{\tilde{\vec{k}} \cdot \tilde{\vec{g}}} \tilde{F}_{s'}\p{\vec{r}, \vec{v}'}, 
    \label{Iss}
\end{equation}
where $\vec{g} = \vec{v} - \vec{v}'$, $d$ is the number of resolved spatial
dimensions in the simulation, $W_s$ is the macroparticle
weight defined as $W_s = N_{\text{D}s}/N_{\text{D}s}^{\text{M}}$
(a species-dependent version of Eq.~(\ref{eq:weight}))
where $N_{\text{D}s} = n_s \lambda_{\text{D}s}^3$ and
$N_{\text{D}s}^{\text{M}} = n_s^{\text{M}} \lambda_{\text{D}s}^d$.
It is assumed here that all macroparticles of the same species carry the same
weight $W_s$. The tilde quantities are normalized
such that $\tilde{\vec{k}} = \vec{k} \lambda_{\text{D}s '}$,
$\tilde{\vec{v}}' = \vec{v}' / \bar{v}_{s'}$, 
$\tilde{\vec{v}} = \vec{v} / \bar{v}_{s'}$, 
$\tilde{\vec{g}} = \p{\vec{v}_{s} - \vec{v}_{s'}} / \bar{v}_{s '}$,
$\tilde{\vec{K}}_{\text{S}} = \vec{K}_{\text{S}}/\lambda_{\text{D}s '}$,
and $\tilde{F}_{s} = \bar{v}_{s}^{d} F_{s}$. $F_{s}\p{\vec{v}} = f_{s}/n_{s}$
is the normalized velocity distribution function.

Here the density, $n_{s}$, describes the physical density such that the
macroparticle density, charge, and mass are
$n_{s}^{\text{M}} = \lambda_{\text{D}s}^{3-d}n_{s}/W_{s}$,
$q_{s}^{\text{M}} = \lambda_{\text{D}s}^{d-3} W_{s} q_{s}$, and
$m_{s}^{M} = \lambda_{\text{Ds}}^{d-3} W_{s} m_{s}$ respectively. In the
reduced dimensions, $q^{\text{M}}$ and $m^{\text{M}}$, are areal charge
and mass density in 1D and lineal in 2D, leading to the common
interpretation that these simulations describe charged surfaces or charged rods,
respectively. 
The thermal speed for each species $s$ is defined
$\bar{v}_s = \omega_{ps} \lambda_{\text{D}s} = \sqrt{k_\text{B}T_s/m_s}$ 
where $\omega_{ps} = \sqrt{n_{s} q_{s}^{2} / \epsilon_{0} m_{s}}$, and
$\lambda_\mathrm{D}^{-2}=\sum_{s}^{}\lambda_{\text{D}s}^{-2}$,
$\lambda_\mathrm{\text{D}s}=\sqrt{\epsilon_{0} k_\mathrm{B}T_{s} / n_{s} q^{2}_{s}}$
is the Debye length. Notably, the Debye length and plasma frequency are
unchanged in the physical and numerical systems alike since they do not depend
on the particle weight or dimensionality. Though the distribution describes the
statistical evolution of macroparticle centers, it is normalized to the
physical density, $\int_{}^{} \diff^{d} v\ f_{s}\p{\vec{r}, \vec{v}} =
n_{s}\p{\vec{r}}$. Additional factors in Eq.~(\ref{eq:A_D_coeff}) contain
species ratios
\begin{equation}
    R_{ss '}^{D} = \frac{1}{2} \frac{q_{s}^{2}}{q_{s '}^2} \frac{m_{s '}^{2}}{m_{s}^{2}}, \qquad 
    R_{ss '}^{A} = \frac{1}{2} \frac{q_{s}^{2}}{q_{s '}^2} \frac{m_{s '}^{2}}{m_{s}^{2}} \p{1 + \frac{m_{s} W_{s}}{m_{s '} W_{s '}}}. 
    \label{species_ratios}
\end{equation}

The integrand of Eq.~\eqref{Iss} expresses how particles exchange momentum by
emitting and absorbing resonant fluctuations. One may alternatively view
this as a Landau~\cite{landau} operator where the colliding particles scatter
via dynamically screened Coulomb potentials.~\cite{balescu} The critical
component of this equation is the screened force kernel
\begin{equation}
    \vec{K}_{\text{S}}\p{\vec{k}} = \vec{K}\p{\vec{k}}/\abs{\varepsilon\p{\vec{k}, \vec{v} \cdot \vec{k}}}. 
\end{equation}
It includes charge screening via the longitudinal plasma dielectric function
\begin{equation}
    \varepsilon \p{\vec{k} , \omega} = 1 + i\vec{K} \cdot \sum_{s}^{} \omega_{ps}^{2}\ \int_{}^{} \pdv{F_{s}}{\vec{v}} \frac{\diff^d v}{ \omega - \vec{k} \cdot \vec{v}}. 
    \label{dielectric}
\end{equation}
The bare force kernel $\vec{K}$ contributes directly to the collision process
of Eq.~\eqref{Iss} as well as to the dielectric response.

Because of its similarities to the traditional plasma theory, the
kinetic equation presented here illuminates some interesting facts about
numerical thermalization in PIC. As shown in Fig.~\ref{fig:regimes}, the
shape and weight effects might conspire to cancel in the collision operator,
but they cannot in the Vlasov mean field term (Eq.~\eqref{mean_field}), which
only depends on $\Delta$ (or $R$). Importantly, all the mass, charge, and weight
ratios factor out of the integral, and the dielectric is symmetric under
species inversion $s\leftrightarrow s'$. Consequently, the ratio of numerical
and physical thermalization times for each $s,s '$ collision pair is
independent of the species mass, charge, and weight ratios, and it depends only
on $W_{s'}$ and $R$. If numerical thermalization is a problem
for electron-electron collisions, it will be a problem for the other
Coulomb collision processes as well (electron-ion and ion-ion), presuming they
are relevant to the dynamics of the physical system. The comparison of the
numerical and physical collision times in a one-component system discussed
here, including Figs.~\ref{fig:regimes} and \ref{fig:tau}, then
trivially extends to multicomponent systems because the ratio
$\tau_\textrm{num}/\tau_\textrm{phys}$ is essentially species independent
(aside from minor modifications of the screening length in multicomponent
systems). 

At this stage, the analysis could be extended to more complex PIC
algorithms. If a simulation uses an unstructured mesh, these forms for the
force kernel
could be applied to a particular location in the simulation with a local
characteristic cell size, provided the cell sizes do not vary rapidly over several
Debye lengths. If a simulation uses dynamic re-weighting and each
particle has its own weight, then a renormalized local weighted velocity
distribution can be used in place of $F_{s}(\vec{v})$ in Eqs.~\eqref{Iss}
and~\eqref{dielectric}.

Adaptive grid and dynamic re-weighting
methods have become common in recent practice.
Making the thermalization time equal to the physical time everywhere is
difficult in these complex simulations that physical have constraints on the grid.
Without a model for numerical thermalization,
these simulations must coarsely reduce thermalization to be negligible
and use Monte Carlo Coulomb collision methods to recover collisional behavior,
potentially wasting resources.

\subsection{Approximations for the drag coefficient}

In this section, successive simplifications to the previous kinetic equations
are discussed in an effort to extend the comparison between PIC and physical
kinetic theory and supply users with practical means of estimating
numerical thermalization in their simulations.

Though the mesh structure and interpolation scheme of PIC calculations can vary
considerably across applications, it is illustrative to focus on the isotropic
shapes as done by Langdon~\cite{langdonJOCP1970} and Okuda~\cite{okudaPOF1970}.
This approximation is made by replacing the true shape of the macroparticle
with an isotropic analogue. For example, the product B-splines of
Eq.~\eqref{bsplines} have analogous isotropic forms that can be constructed
from convolutions of the uniform $n$-D ball distribution of radius $R$ and unit
volume. The Fourier transforms of these splines take the following form:
\begin{equation}
    S_{\text{iso}}^{\p{n}}\p{k, R} =
    \begin{cases}
        \p{\frac{\sin\p{k R}}{k R}}^{n+1} & \text{ in 1D,} \\
        \p{2\frac{J_{1} \p{k R}}{k R}}^{n+1} & \text{ in 2D,} \\
        \p{3\frac{\sin \p{k R} - k R \cos\p{kR}}{\p{kR}^{3}}}^{n+1} & \text{ in 3D,} 
    \end{cases}
    \label{isotropic_shapes}
\end{equation}
where $R$ can be set to the half-width of the mesh spacing, $\Delta/2$, or the
half-width of the filter, $R=n_\mathrm{f} \Delta/2$, if a filter matching
Eq.~\eqref{ngp_filter} is used. Here, $J_1$ is the first-order Bessel function
of the first kind. 

Just as the $n$-spheres (and their higher-order convolutions) are isotropic
$S\p{\vec{r}} \to S\p{r}$, so too are their Fourier transforms
$S\p{\vec{k}} \to S\p{k}$. Therefore when $\Delta \lesssim \lambda_\mathrm{D}$
and filters are approximately isotropic, $\varphi\p{\vec{k}} \to \varphi\p{k}$,
then $\mathcal{D} \approx 1$ and all the PIC modifications to the bare force
kernel are also isotropic
\begin{equation}
    \vec{K}(\vec{k}, R) = -\frac{i\vec{k}}{k^2}P(\vec{k}, R), \quad P\p{\vec{k}, R} \to P\p{k, R} \equiv  S^{2} \varphi^{2} 
\end{equation}
and depend only on the radius of the macroparticle $R$.

Combining isotropic shapes
with Maxwellian field distributions
$F_{s'} = (2 \pi \bar{v}_{s '}^2)^{-d/2} e^{-v^2/2 \bar{v}_{s'}^2}$ permits a
double-integral representation of Eq.~\eqref{Iss}. For example,
the normalized drag integral defined
$\vec{I}_{A} \equiv \partial_{\vec{v}} \cdot \mathbb{I}_{ss'}$
has magnitude
\begin{equation}
    I_{A}^{\text{dyn}} =  \tilde{v}\ W_{s'}
    \int_{0}^{\pi} \diff \theta \int_{0}^{\infty} \diff \tilde{k}
    \renewcommand\arraystretch{1.5}
    \begin{Bmatrix}
    \sqrt{\frac{2}{\pi}} \delta\p{\theta} \\
    \frac{1}{\sqrt{2} \pi^{3/2}} \\
    \frac{\sin \theta}{\p{2 \pi}^{3/2}}  \\
    \end{Bmatrix}_{d}
    \frac{\tilde{k}^{d-4} P^{2} \cos^{2} \theta }{\abs{\varepsilon\p{\tilde{k}, \tilde{v}, \theta}}^{2}}
    e^{-\frac{\tilde{v}^{2}}{2} \cos^{2} \theta}.
    \label{Iss_dyn}
\end{equation}
It is proportional to the friction
force density acting antiparallel on a beam of particles $s$ streaming through
a Maxwellian distribution of particles $s'$ at velocity $v$. 
The $\cu{\cdot}_d$ notation denotes the different dimensions ($d = 1,2,3$)
from top to bottom. The angle is defined by
$\vec{v} \cdot \vec{k} \equiv vk\cos \theta$. It is to be understood that the
$\theta$ integral is over both sides of the $\delta$ function in the 1D case.
Upon substitution $d \to 3$, the drag given by Okuda~\cite{okudaPOF1970}
is recovered. Equation~\eqref{Iss_dyn} retains dynamic screening effects
expressed through the dielectric dependence on the velocity $\vec{v}$ and its angle
relative to the wave vector $\vec{k}$.
\begin{figure}
  \centering
  \includegraphics[width=\linewidth]{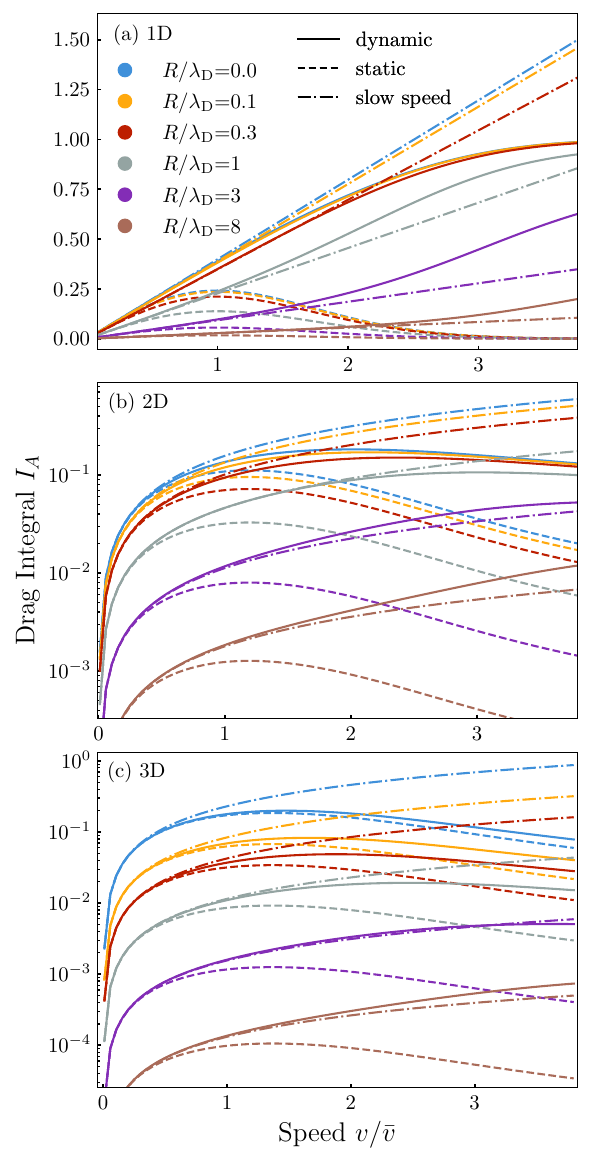}
  \caption{The drag coefficient prediction for 1D-3D dimensional one-component simulations with various
  macroparticle shape radii and weight $W = 1$. The dynamic (Eq.~\eqref{Iss_dyn}),
  static (Eq.~\eqref{Iss_stat}), and slow speed (Eq.~\eqref{Iss_slow})
  models are compared. The zero-order (top hat) shape function of Eq.~\eqref{isotropic_shapes}
  with radius $R$ was used in all cases.}
  \label{fig:drag}
\end{figure}

At lower speeds, $v \lesssim \bar{v}_{s'}$, the dielectric response of the
plasma around each particle can be approximated as a static distribution
(independent of $v$) with the replacement
$\varepsilon\p{\vec{k} , \vec{v} \cdot \vec{k}} \to \varepsilon\p{\vec{k}, 0}$.
In this case, the collision integral becomes separable
\begin{equation}
    I_{A}^{\text{static}} = \tilde{v}\ W_{s'} I_{\theta} \p{v} I_{S},
    \label{Iss_stat}
\end{equation}
where the angular integral is defined as
\begin{equation}
    I_{\theta}\p{v} = \int_{0}^{\pi} \diff \theta
    \renewcommand\arraystretch{1.5}
    \begin{Bmatrix}
    \sqrt{\frac{2}{\pi}} \delta\p{\theta} \\
    \frac{1}{\sqrt{2} \pi^{3/2}} \\
    \frac{\sin \theta}{\p{2 \pi}^{3/2}}  \\
    \end{Bmatrix}_{d}
    \cos^{2} \theta e^{-\tilde{v}^{2} \cos^{2} \theta / 2}.
\end{equation}
Evaluating this provides
\begin{equation}
I_\theta\p{v} = 
    \begin{cases}
        \sqrt{\frac{2}{\pi}}e^{-\tilde{v}^{2}/2}, & \text{in 1D} \\
        \frac{1}{2\sqrt{2 \pi}}e^{-\tilde{v}^2/4} \left[I_0(\frac{\tilde{v}^2}{4})-I_1(\frac{\tilde{v}^2}{4})\right], & \text{in 2D} \\
        \frac{\psi\p{\tilde{v}^{2}/2}}{2 \pi\tilde{v}^{3}}, & \text{in 3D} , \\
    \end{cases}
\end{equation}
which corresponds to the usual Maxwellian stopping power expression related to
the Maxwell integral in 3D,
\begin{equation}
    \psi\p{x} \equiv \frac{2}{\sqrt{\pi}} \int_{0}^{x} \diff t\ t^{1/2} e^{-t} = \text{erf} \p{\sqrt{x}} - \frac{2}{\sqrt{\pi}} \sqrt{x} e^{-x},
\end{equation}
and the modified Bessel functions of the second kind, $I_0, I_1$, in 2D.

In Eq.~(\ref{Iss_stat}), the shape integral is independent of speed and
is defined as
\begin{equation}
    I_{\rm S} = \int_{0}^{\infty} \frac{\tilde{k}^{d-4} P^{2}}{(1+P/\tilde{k}^2)^2}\diff \tilde{k}.
    \label{eq:shape}
\end{equation}
In 3D, $I_{\rm S}$ was identified by Langdon~\cite{langdonPOF1970} as
converging to the Coulomb logarithm
($I_{\rm S} \rightarrow \ln \Lambda_{ss^\prime}$) in the physical limit of
point particles ($P \to 1$) if the upper bound is truncated at the plasma
parameter $\Lambda_{ss '}$. This corresponds to the traditional~\cite{balescu}
choice of ignoring wavenumbers above $\tilde{k} = \Lambda_{ss '}$ where
$\Lambda_{ss'} = \lambda_\text{D}/b_{ss'}$,
$b_{\text ss'} = q_s q_{s '} / (4 \pi \epsilon_0 m_{ss'} \bar{v}_{ss'}^2)$ is the
thermal distance of closest approach,
$m_{ss'} = m_s m_{s '}/(m_s + m_{s '})$ is the reduced mass,
and $\bar{v}_{ss'}^2 = \bar{v}_s^2 + \bar{v}_{s '}^{2}$ is the combined thermal speed.
As Okuda~\cite{okudaPOF1970} has pointed out, no truncation is
required in 1D or 2D because the integral converges, even for point particles
$(P\rightarrow 1)$. For finite-sized particles, $R > 0$, Eq.~\eqref{eq:shape}
converges in all dimensions and no truncation is required. In principle, a
shape function with a radius $R \ll b_{\rm ss'}$ would require a cutoff for the
same reason that point particles do. However, it is generally expected that
$R \gtrsim b_{ss'}$ in electrostatic PIC simulations, since a PPPM molecular
dynamics~\cite{eastwood,hockney} simulation would be more efficient at
resolving the close-ranged interactions than using a grid that resolves
$b_{ss'}$. The results in Fig.~\ref{fig:tau} are therefore insensitive to the
cutoff, provided it is at or above~$\Lambda_{ss'}$.

As a final approximation, the slow speed limit of the drag integral may be
easily derived from Eq.~\eqref{Iss_stat}. The result is a drag integral
proportional to the speed
\begin{equation}
    I_A^{\rm slow} =
    \left\{\begin{array}{@{} c @{}}
    \sqrt{\frac{2}{\pi}}           \\
    \frac{1}{2\sqrt{2 \pi}}       \\
    \frac{1}{3 \sqrt{2 \pi^{3}}}
  \end{array}\right\}_{d} \tilde{v}\ W_{s'} I_{S}. 
  \label{Iss_slow}
\end{equation}
Fig.~\ref{fig:drag} shows the
magnitude of $I_A$ in the dynamic, static, and slow limits for a range of
macroparticle radii and dimensions. The results of
Eqs.~\eqref{Iss_dyn} and \eqref{Iss_stat} are in agreement with
Okuda,~\cite{okudaPOF1970} however shown here are results for the top hat shape
$S^{(0)}$ rather than the Gaussian shape $S(k) = e^{-k^2 R^2}$. As anticipated,
the magnitude of the drag decreases monotonically for wider shape functions at
all speeds. The static screening approximation is found to be increasingly poor
at faster speeds and reduced dimensions for the top hat shape as well.
Interestingly, the dynamic and slow speed models agree well at intermediate
speeds ($v~\approx~\bar{v}$) and intermediate macroparticle sizes
($R \approx \lambda_{\rm D}$), whereas the static approximation only agrees at
slow speeds. It is also noteworthy that the Bragg peak shifts to higher speeds,
particularly in the most complete ``dynamic'' model, as the particle width
increases beyond the Debye length.

\subsection{PIC Numerical collision time}
The relationship between the velocity-dependent Fokker-Planck coefficients and
the corresponding transport coefficients is provided by velocity moments taken
over the plasma kinetic equation in the usual way.~\cite{grad1949,grad1958}  As an example
of the influence of shape effects on numerical collisions, the friction force
density between two drifting Maxwellian distributions $f_s$ and $f_{s '}$ is
obtained from the first velocity moment of the collision operator (RHS of
Eq.~\eqref{fokker_planck}). In the limit of a slow drift speed
$V_{s '} \ll \bar{v}_{ss '}$, the friction force density on each distribution is
proportional to the drift
$\vec{F}^{\text{frc}}_{ss '} = -n_{s} m_s \nu_{ss '} \p{\vec{V}_{s}-\vec{V}_{s '}}$.
This assumption corresponds to the linear constitutive relation assumed in the
first-order Chapman-Enskog solution to the friction force coefficient.
In this fluid description of two drifting populations, both distributions will
drag toward each other simultaneously until the drift
$\Delta \vec{V}_{ss '} = \vec{V}_s - \vec{V}_{s'}$ is eliminated. For small
flow shifts $\abs{\Delta V_{ss '}} \ll \bar{v}_{ss'}$, this overall
relaxation is described by
\begin{equation}
    \dvOne{t} \Delta \vec{V}_{ss '} = - \nu_{\text{V}ss '} \Delta \vec{V}_{ss'},\qquad 
    \nu_{\text{V}ss'} \equiv \nu_{ss '} + \nu_{s's} 
\end{equation}
such that the multispecies collision time is defined as
\begin{equation}
    \tau_{ss'} \equiv \frac{1}{\nu_{\text{V}ss'}}.
\end{equation}
The collision rate $\nu_{ss'}$ associated with this drift processes is
given by
\begin{equation}
    \nu_{ss'}
    = \frac{\omega_{ps'} R_{ss '}^{A}}{N_{\text{D}s '}}
    \frac{\bar{v}_{s '}}{\bar{v}_{s}}
    \renewcommand\arraystretch{1.5}
    \begin{Bmatrix}
        \sqrt{\frac{2}{\pi}} \\
        \frac{1}{2} \\
        \frac{\sqrt{2}}{3 \sqrt{ \pi}} \\
    \end{Bmatrix}_d
    \int_{0}^{\infty} \frac{\diff v}{\bar{v}_{s}}\ 
    \frac{v^{d}}{\bar{v}_{s}^{d}} I_{A}\p{v} e^{-v^{2}/2\bar{v}_{s}^{2}} .
    \label{v_integral}
\end{equation}

This model for the collision time applies only for distributions with
small drift speeds. When modeling the collision time for large drifts or electron
tails, retaining the velocity dependence of the distribution is necessary for
taking moments of the drag integral, $I_\mathrm{A}\p{v}$. As shown
in Fig.~\ref{fig:drag},
the dynamic drag model depends on macroparticle size and speed in a non-separable
way.
For distributions that are in a linear relaxation regime, collision
time models for each the dynamic, static, and slow speed drag integrals
can be evaluated to estimate numerical thermalization with varying ease.
Inserting the slow speed approximation into Eq.~\eqref{v_integral} grants a collision rate proportional
to the stopping power slope for slow speeds $v \to 0$,
\begin{equation}
    \label{nu_slow}
    \nu_{ss'}^{\text{slow}} = \frac{\omega_{ps '} R_{ss '}^{A}}{N_{\text{D}s '}} 
    \renewcommand\arraystretch{1.5}
    \left\{\begin{matrix}
    \sqrt{\frac{2}{\pi}}           \\
    \frac{1}{2\sqrt{2 \pi}}       \\
    \frac{1}{3 \sqrt{2 \pi^{3}}}
    \end{matrix}\right\}_{d}
    W_{s'} I_{S}.
\end{equation}
Inserting the static screening approximation grants the usual Coulomb collision
rate solution multiplied by the weight factor $W_{s '}$ and with the Coulomb
log replaced with $\ln\Lambda_{ss '}~\to~I_\text{S}$,
\begin{equation}
    \label{nu_static}
    \nu_{ss'}^{\text{static}} =
    \begin{Bmatrix}
    \vspace{1mm}\frac{1}{2\sqrt{\pi}} \\
    \vspace{1mm}\frac{\sqrt{\pi}}{8} \\
    \vspace{1mm}\frac{1}{12\pi^{3/2}}  \\ 
    \end{Bmatrix}_{d}
    \p{\frac{2}{1 + \p{\bar{v}_{s} / \bar{v}_{s '}}^2}}^{3/2}
    \frac{\omega_{ps'} R_{ss '}^{A}}{N_{\text{D}s'}}
    W_{s '}I_{\rm S}.
\end{equation}
Though this is only one kind of relaxation, other charged particle
thermalization processes (e.g. velocity diffusion) have the same character
regarding macroparticle shape effects. This fact is demonstrated by the
general agreement between each of the drag models in Fig.~\ref{fig:tau}
and the separability of $P$ and $W_s$ from the
otherwise physical drag integral in the static screening approximation.

Combining Eqs.~\eqref{Iss_dyn}, and~\eqref{v_integral}
for a Maxwellian $s$ distribution grants the dynamic screening expression
for the Coulomb collision frequency
\begin{widetext}
\begin{equation}
    \nu_{ss'}^{\text{dyn}}= \omega_{ps'} \frac{W_{s'}R_{ss '}^{A}}{N_{\text{D}s '}}
    \int_{0}^{\infty} \frac{\diff v}{\bar{v}_{s}}
    \p{\frac{v}{\bar{v}_{s}}}^{d+1}
    e^{-v^{2}/2\bar{v}_{s}^{2}}
    \int_{0}^{\pi} \diff \theta
    \renewcommand\arraystretch{1.5}
    \begin{Bmatrix}
    \frac{2}{\pi} \delta\p{\theta} \\
    \frac{1}{\p{2 \pi}^{3/2}} \\
    \frac{\sin \theta}{6 \pi^{2}}  \\
    \end{Bmatrix}_{d}
    \cos^{2} \p{\theta}
    e^{-\frac{v^{2} \cos^{2} \theta}{2 \bar{v}_{s '}^{2}}}
    \int_{0}^{\tilde{k}_\mathrm{max}} \diff \tilde{k}
    \frac{\tilde{k}^{d-4} P^{2}}{\abs{\varepsilon\p{\tilde{k}, \tilde{v}, \theta}}^{2}}. 
    \label{big_nu}
\end{equation}
\end{widetext}
Equation~\eqref{big_nu} was evaluated using numerical quadrature to
produce the main theoretical collision time results in
Figs.~\ref{fig:regimes} and~\ref{fig:tau} alongside Eqs.~\eqref{nu_static}
and~\eqref{nu_slow} for the static and slow speed
approximations. The contours of $\tau^{\text{num}}/\tau^{\text{phys}}(W, R)$
in Fig.~\ref{fig:regimes} were evaluated from the dynamic model
for a single species,  $S = S^{\p{0}}$ case. The physical limit
$\tau^{\text{phys}}$ was obtained by allowing the macroparticle radius to
go to $0$ in Eq.~\eqref{big_nu}. A cutoff of
$\tilde{k}_\mathrm{max} = 120 \pi$ was used for both physical and numerical
3D cases.

The analytical form of this result provides important information about how
the drag process scales in multispecies simulations. In particular, the mass
ratio scaling laws for the physical momentum transfer times, namely
$\tau_\mathrm{ee} \propto \tau_\mathrm{ei} \propto \sqrt{m_\mathrm{e}/m_\mathrm{i}} \tau_\mathrm{ii}$,
also hold for numerical collisions. This is especially relevant to PIC
simulations with both kinetic electrons and ions. These simulations must be
long enough to resolve ion dynamics which occur roughly
$\sqrt{m_\mathrm{i} / m_\mathrm{e}}$ times slower than the electron dynamics.
However, the simulated electrons still thermalize within approximately
the electron collision time, $\tau_{ee}$. Since PIC is only a good
approximation to the Vlasov equation on timescales $t \ll \tau_{ss'}$,
comparatively longer simulations that track the dynamics of  heavier ion species will encounter
more numerical electron thermalization. This may also influence the
computational work required of the simulation:
in order to maintain the same numerical thermalization ratio
$t_\text{sim} \nu_{\rm Vee}$ as in the single component case, the runtime
complexity can scale on the order
$\mathcal{O}(N N_{t}) \sim \mathcal{O}(m_\mathrm{i}/m_\mathrm{e})$. The number of time steps,
$N_{t}$, will increase by roughly $\sqrt{m_\mathrm{i}/m_\mathrm{e}}$ in order to resolve
the ion dynamics. The number of macroparticles, $N$, must then increase by the
same factor to prevent numerical thermalization. If numerical thermalization
were not accounted for, the runtime complexity would instead be on the order
of $\mathcal{O} \fp{\sqrt{m_\mathrm{i}/m_\mathrm{e}}}$
because of the time step increase alone.

\begin{figure*}
  \centering
  \includegraphics[width=\linewidth*6/7]{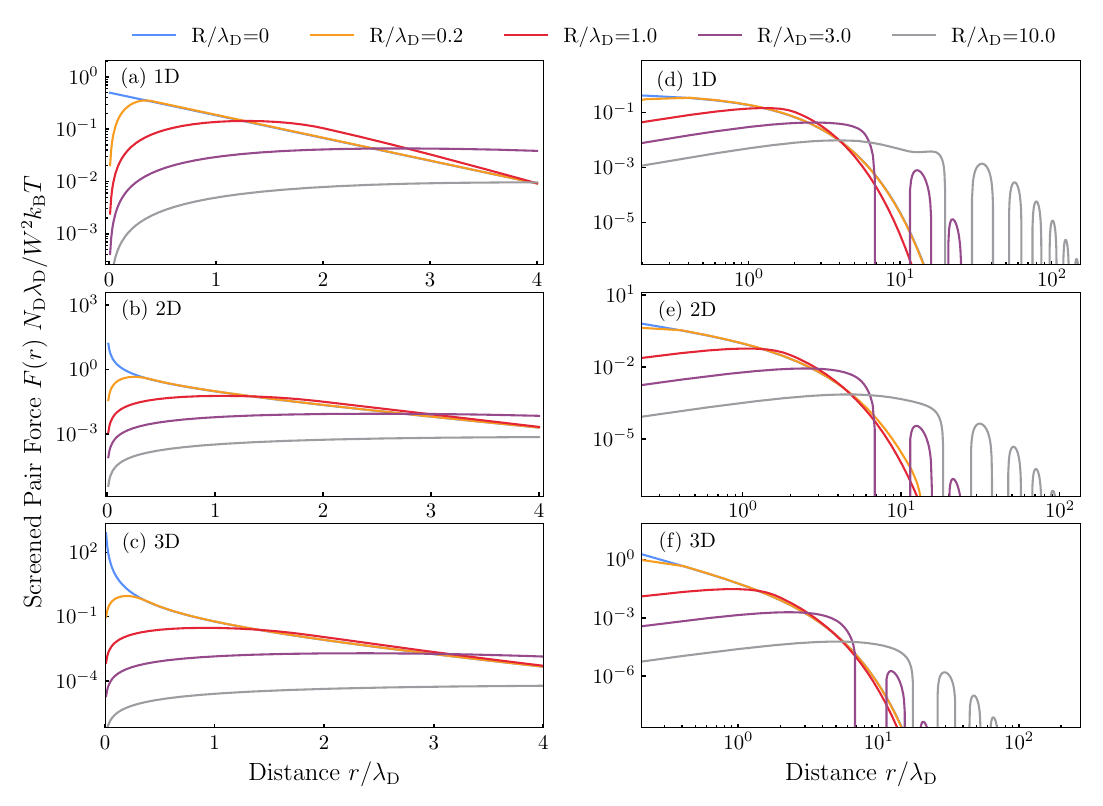}
  \caption{The static screening of the pair force between identical isotropic
  $S^{(0)}$ macroparticles. For large shapes, the screened force law oscillates
  at larger distances (panels d-f).}
  \label{fig:force}
\end{figure*}

It may be possible to alter the macroparticle shapes via filters or grids in
order to reduce computation time, however this approach has its own
limitations. All the Vlasov terms in the kinetic equation depend only on shape
effects and not the particle weight. Therefore the wave dispersion,
damping,~\cite{langdonPOF1970} charge screening,~\cite{okudaPOF1970} and mean
field of the simulation will differ from their physical values at longer length
scales if the shapes are made large to compensate for a large weight. That is,
in the regime of $R \gg \lambda_{\text D}$, shape effects significantly
improve numerical thermalization at the expense of accuracy in the
remaining collisionless processes. These errors are illustrated by the static
screened force depicted in Fig.~\ref{fig:force} evaluated
numerically by
\begin{subequations}\begin{align}
    \vec{F}(\vec{r}) &= W^{2} \frac{k_\mathrm{B}T}{N_\mathrm{D} \lambda_\mathrm{D}}\mathcal{F}^{-1}
    \frac{- i \vec{k}}{k^2} \frac{P(\vec{k})}{\epsilon(\vec{k}, 0)}
    \mathcal{F} \delta(\vec{r}'), \\
    \frac{F\p{r} N_\mathrm{D} \lambda_\mathrm{D}}{W^{2} k_\mathrm{B}T}
    &= -\int_{}^{} \frac{\diff^{d} \tilde{k}}{\p{2 \pi}^{d}}\
    \frac{\cos \chi}{\tilde{k}} \frac{iP}{1 + P/\tilde{k}^{2} } e^{i \tilde{k} \tilde{r}\cos \chi }
\end{align}\end{subequations}
where
$\vec{r} \cdot \vec{k} \equiv rk\cos \chi$ and $\tilde{r} = r/\lambda_{\rm D}$.
It represents the force field surrounding a macroparticle at the origin due its
own charge distribution, $S^{(0)}$, as well as the dielectric response of the
plasma to the macroparticle. As the macroparticle shape is widened, the
screened force is weakened significantly at short range. As seen from
Eq.~\eqref{Iss}, this force contributes directly to the first-order collisional
processes predicted by the theory. When the macroparticle radius is much less than
the Debye length, the screened force matches the point particle limit for
distances at and longer than the radius. This corresponds to the weak
dependence of numerical thermalization on the macroparticle size in the
$R \lesssim \lambda_\mathrm{D}$ regime shown by Figs.~\ref{fig:regimes}
and~\ref{fig:tau}. When the macroparticle radius is much larger than the Debye
length, the numerical thermalization time is significantly extended, but the
screened force exhibits dramatic errors that extend beyond the macroparticle
radius. These oscillations in the force produce roots in the plasma
dispersion relation~\cite{langdonPOF1970} and peaks in the collision operator
integrand. These errors are usually addressed for each application
(or each characteristic region of a simulation) in order to determine a maximum
permissible shape width. Once this width is reached, numerical thermalization
must be addressed by decreasing the macroparticle weight.

A similar issue arises in various proposed mitigation techniques. Artificially
increasing~\cite{jubinPOP2024} the permittivity of free space $\varepsilon_{0}$
will increase $N_{\text{D}}^{\text{M}}$ and thereby reduce the collision term
of the kinetic equation. It will also weaken the mean field, $\bar{\vec{E}}$,
given by Eq.~\eqref{mean_field}
and weaken the linear response of the plasma (i.e. the collisionless processes)
uniformly as seen through the reduction of $\omega_{ps}$ in
Eq.~\eqref{dielectric}. Thus, the $\epsilon_0$ scaling technique is only valid
if mean field and wave phenomenon are unimportant for the entire simulation.
Some simulations use self-similarity~\cite{taccogna} arguments that scale the
size of the simulation domain to reduce the computational burden. The
permissible extent of such scaling is limited by numerical thermalization
because the PIC Coulomb collision time is determined by $N_\mathrm{D}^{\text{M}}$
regardless of the $N_\mathrm{D}$ in the hypothetically smaller system. Ideal
Vlasov fluids are self-similar in this way, but PIC simulations are not in
general self-similar because of macroparticle collisions. Simulations
also commonly use reduced mass ratios to save computation
time.~\cite{daughton,fox} Depending on the decrease in simulation time
in the reduced mass case and the type of interacting particles, the ratio
$t_\mathrm{sim}/\tau^{\text{num}}$ can possibly increase.

\subsection{Higher-order shape functions\label{sec:shapes}}

\begin{figure*}
  \centering
  \includegraphics[width=\linewidth*6/7]{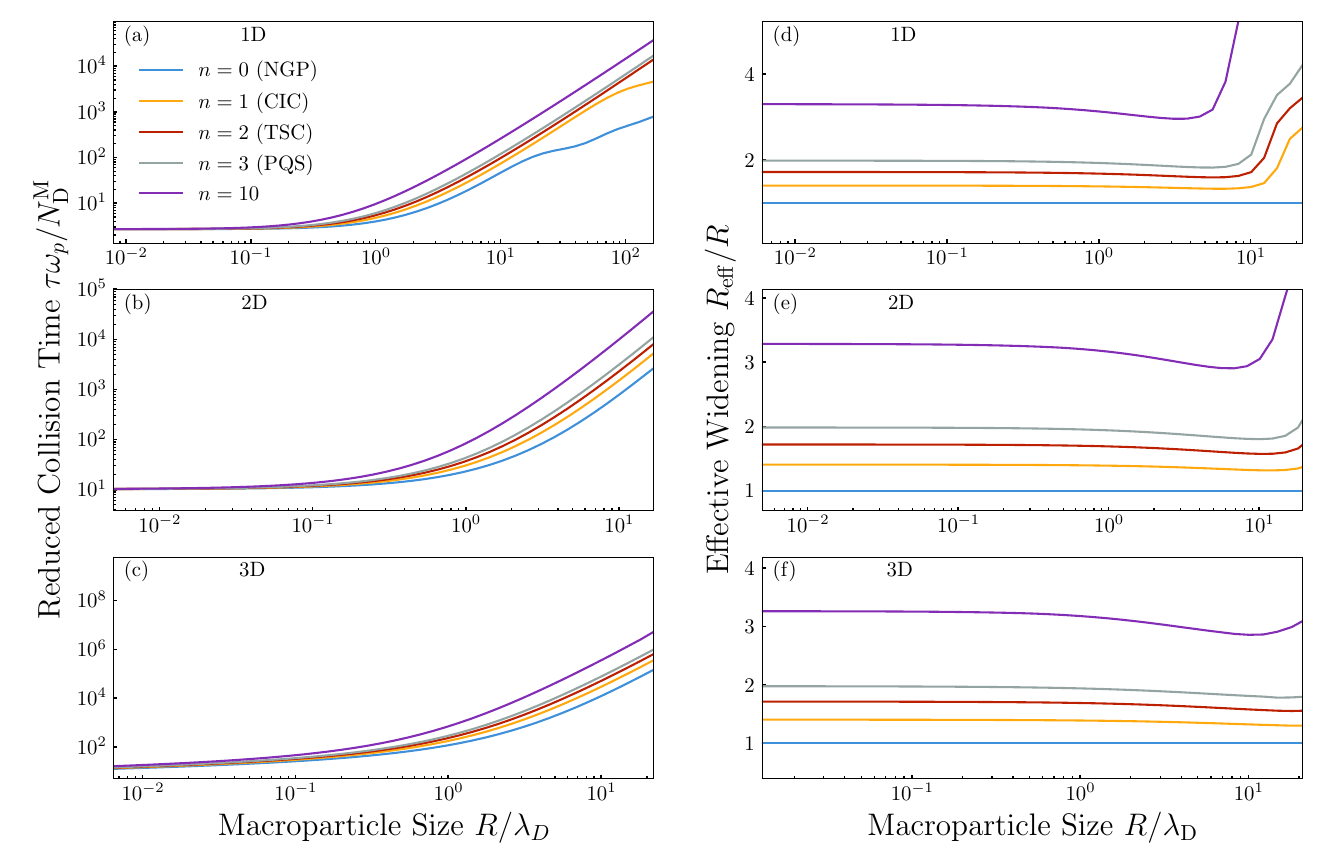}
  \caption{Collision time predictions for the isotropic higher-order shape functions $S^{(n)}$ (a-c). The effective radius is defined such that
  $\tau[S^{(0)}(R_{\rm eff}^n)] \equiv \tau[S^{(n)}(R)]$
  can be computed in
  analogy with the NGP case (d-f).}
  \label{fig:high_order}
\end{figure*}

The higher-order shape functions, $S^{\p{n}}\p{\vec{r}}$, are commonly
used to improve grid heating and reduce noise at the expense of using
higher memory bandwidth to access more cells per macroparticle.
Their isotropic counterparts
$S_{\text{iso}}^{\p{n}}\p{\vec{r}}$ can be readily evaluated by
Eq.~\eqref{big_nu}. Figure~\ref{fig:high_order}
shows the near-equilibrium numerical collision time for several
orders of shape functions. Since the qualitative behavior is seen to be
similar between orders, the results can be inverted to demonstrate that
when $R \lesssim 5\lambda_\mathrm{D}$,
an increase in the spline order grants the same thermalization time
as instead increasing the shape width.
The thermalization time can be expressed through its dependence
on the shape of order $n$ and width $R \equiv n_{\rm f} \Delta/2$ as $\tau [S^{(n)}(R)]$.
The effective size $R_{\text{eff}}$ is defined by
$\tau [S^{(0)} (R^{n}_{\text{eff}})] \equiv \tau[S^{(n)}(R)]$.
Panels (d-f) display the effective widening of the particle
size through the ratio $R^n_{\text{eff}}/R$.
For unfiltered simulations ($n_{\text{f}} = 1$), this numerical thermalization model is only valid for $R = \Delta /2 \lesssim \lambda_{\rm D}$, where $R_{\rm eff}/R$ is constant.
For example, an unfiltered simulation with cubic spline interpolation
($n=3$) and $\Delta = 0.2 \lambda_\mathrm{D}$ will have the
same characteristic collision time as a simulation with nearest grid point ($n=0$)
interpolation and $\Delta = 0.4 \lambda_\mathrm{D}$.
The previous results can therefore account for higher
order shapes by simply multiplying the macroparticle size by
$R_{\rm eff}/R$ when the Debye length is resolved.
In the large $R \gg \lambda_\mathrm{D}$
limit, the lower shape orders reach different asymptotic behavior, but once
$n \gtrsim 2$, the marginal effect of raising the order steadily decreases.
The net takeaway is that the reduction in numerical thermalization from using
higher order shape functions is simply associated with a larger effective
particle width. 

\section{Discussion}
\label{application}

\begin{table*}
\caption{Approximate comparison of physical and numerical thermalization times
in various particle-in-cell simulations. All collision times refer to
electron-electron collisions and linear $S^{\p{1}}$ shape functions.}
\label{tab:tau_compare}
\centering
\begin{ruledtabular}
\begin{tabular}{cccccccccc}
System & Dimension & $L/\lambda_\text{D}$ & $R/\lambda_\mathrm{D}$ & $\tau^\mathrm{phys} \omega_{p}$ & $\tau^\mathrm{num} \omega_{p}$ & $t_\mathrm{sim} \omega_{p}$ & $N_\mathrm{D}/N_\mathrm{D}^{\text{M}}$ & $I_{\rm S}^{\text{phys}} / I_{\rm S}^{\rm num}$ & $\tau^\mathrm{phys}/\tau^\mathrm{num}$ \\
\hline
ICF Hotspot~\cite{van_de_wetering}                      & 1 & 6e+7 & 1e+3 & 5e+0 & 5e+4 & 3e+6 & 2e+0 & 2e+4 & 9e-5 \\
Magnetic Reconnection~\cite{fox,nilson}                 & 2 & 8e+3 & 5e-1 & 5e+5 & 2e+3 & 3e+4 & 6e+2 & 2e+0 & 3e+2 \\
Penning Discharge~\cite{fubiani}                        & 2 & 2e+2 & 3e-1 & 8e+4 & 4e+3 & 3e+5 & 4e+1 & 2e+0 & 2e+1 \\
Hall Thruster~\cite{croes}                              & 3 & 6e+2 & 5e-1 & 7e+4 & 8e+3 & 2e+4 & 4e+2 & 4e+1 & 8e+0 \\
Streamer Discharge~\cite{teunissen}                     & 3 & 3e+2 & 5e+1 & 6e+2 & 2e+2 & 1e+4 & 6e+6 & 3e+6 & 3e+0 \\
ITER Neutral Beam Injector~\cite{wunderlich,mochalskyy} & 3 & 2e+3 & 1e+1 & 5e+3 & 4e+2 & 2e+7 & 2e+5 & 1e+4 & 1e+1 \\
\end{tabular}
\end{ruledtabular}
\end{table*}

The validated kinetic theory is now applied to discuss ways in which modern PIC simulations
should consider, and may be influenced by, numerical thermalization. The broad goal in a kinetic
simulation is to accurately model each distribution function $f_s$ and its moments over
a particular spatial and temporal scale. 
If collisions are expected to influence the physical system, then one must ensure that the
numerical thermalization rate is far below the physical one so that the physical collision rate can be added in using Monte Carlo. 
If collisions are not expected to influence the physical system, then one must check that the, often enhanced, numerical thermalization remains negligible over the time and length scales of interest. 
Quantifying this, the PIC thermalization time (on the order of $\tau_{ss '}^{\text{num}}$) should
therefore be compared to both the physical Coulomb collision time
$\tau_{ss '}^{\text{phys}}$ for each species combination, $s,s'$, as well as to other characteristic
timescales of the simulation.
As discussed in Ref.~\onlinecite{jubinPOP2024}, determining
if numerical thermalization is an issue requires comparing the timescales
for all the relevant thermalizing processes (e.g. electron-neutral
scattering) and the lifetime of each species. As usual, it is also necessary
to compare the macroparticle size to other relevant length scales of the simulation
in order to resolve the important physics.

A small sample of recent studies have been selected to illustrate these considerations 
in a variety of interesting systems. We emphasize that these examples are
selected as representative of common practices in current PIC simulation work.
Any predicted thermalization issues are not particular to these examples, nor
do the they necessarily call into question their principal results. 
They are simply meant to illustrate the types of concerns that ought to be raised by numerical thermalization when a practitioner carries out a PIC simulation. 
Based on Eqs.~\eqref{Iss_dyn}
and~\eqref{v_integral}, and the commonly-used linear
(``cloud-in-cell'') shape function (Eq.~\eqref{isotropic_shapes}
with $n=1$), a coarse estimate of the electron-electron thermalization time is
summarized in Table~\ref{tab:tau_compare} for each simulation.
A more dedicated analysis of
each system could use non-equilibrium distributions, specific shape functions,
and spatially resolved, multispecies collision times from the PIC data directly
in Eq.~\eqref{Iss}. Here, characteristic densities and
temperatures are used and equilibrium distributions are assumed instead. This
approach shall serve as a reasonable first-order estimate of the numerical
thermalization time for many non-uniform and non-equilibrium systems.

We first consider a simulation of a 1D spherical ICF fusion capsule implosion.
This simulation pushes the limits of PIC close to
nearly degenerate and moderately coupled matter. The estimates in
Table~\ref{tab:tau_compare} are based on the initial density and temperature at
the 50 $\mu$m radius. These conditions result in a nearly fluid-like simulation
with a short physical collision time and very small Debye length. Numerical
thermalization is not a problematic constraint here because reaching the
physical $N_{\text{D}}^{} \approx 2$ is feasible and the shape effects extend
the collision time well beyond $\tau_{\text{phys}} \omega_p$. The collisions
are then appropriately implemented via Monte Carlo Coulomb collisions. However,
this is achieved through a very large macroparticle width of
$\sim 1000\lambda_\textrm{D}$, which smooths the inter-particle forces over
very long distances. 

Next, consider a 2D simulation of magnetic reconnection in laser-induced
plasma bubbles~\cite{fox} used to model an experiment.~\cite{nilson}
A principal quantity of interest in this case is the reconnection rate
and how it is influenced by magnetic flux pile up in the current sheet.
In this experimental regime, the reconnection rate is expected to be
set by collisionless processes and not set by classical
resistivity.
The transition between collisional (resistivity-dominated)
and collisionless reconnection is expected to occur when the ion skin
depth ($d_{\rm i} = c/\omega_{p\text{i}}$) is less than the Sweet-Parker
current sheet thickness ($\delta_\textrm{SP} = L/\sqrt{S}$), where
$S=LV_A/\eta$:  $\delta_{\text{SP}} \lesssim d_{\rm i}$. Here $V_\mathrm{A} = B/\sqrt{\mu_{0} m n}$
is the Alv\'en speed and $\eta$ is the electrical resisitivity defined in Ohm's law.
Determining whether or not the simulation captures the collisionless, versus
collisional, regime therefore depends on the numerical collision rate.
In the experiment, the plasma is expected to be in a collisionless regime
because the current sheet width is less than the ion inertial scale
($\delta_\textrm{SP}/ d_i \approx 0.4-0.8$). Since the Sweet-Parker width is
proportional to the square root of resistivity,
$\delta_\textrm{SP} \propto \sqrt{\eta} \propto \sqrt{1/\tau_{ee}}$,
and the simulations did not have an explicit Coulomb collision routine
included, one might expect that the simulation is also in a collisionless
regime at all times. However, because numerical thermalization sets a
minimum resistivity, one should check that this does not influence the expected
reconnection regime. Table~\ref{tab:tau_compare} 
indicates that the numerical resistivity is approximately 300 times the physical
value. This might influence the early time behavior in the simulation before
the dynamics of magnetic flux pileup eventually cause a transition to the
collisionless reconnection regime due to compression heating in the current
sheet. Since the classical resistivity itself sets the width of the
Sweet-Parker current sheet, the present model for the numerical collision time
could be used to inform a maximum macroparticle weight required to simulate
physical scenarios that are strictly in the fast reconnection regime.

As an example of where numerical collisions might be present but not influence
the main expected outcomes of a simulation, consider a 3D PIC simulation of a
partially ionized Penning discharge~\cite{fubiani} used to study the
evolution of rotating azimuthal mode structures. In this physical system, the
electrons are magnetized such that $\omega_{pe} \gg \omega_{ce} \gg \nu_e$
where $\nu_e$ is the total electron collision frequency, including Coulomb
collisions and collisions with neutrals. In this regime, the collision
processes become characterized by conductivity tensors, which have modified
scalings in all directions except parallel to the magnetic field. The
unmagnetized kinetic theory for numerical collisions is expected to work in
this magnetization regime because of the condition that the gyrofrequency is 
small compared to the plasma frequency, $\omega_{ce} \ll \omega_{pe}$. This
simulation reports an electron-neutral collision time of
$\omega_{p\text{e}} \tau_{\rm eN} \approx $ 60 which was implemented with a
Monte Carlo collision routine. Although there is considerable numerical
thermalization, the (real) thermalization of electrons with the neutrals is
dominant on average. This illustrates a situation where the
numerical collision time is artificially much shorter than the physical value,
but where that might not matter for the simulation outcomes because the imposed
electron-neutral collision time is much shorter than the numerical Coulomb
collision time. 

As an example where numerical thermalization may compete with wave-particle
interaction processes, consider 2D PIC simulations of anomalous cross-field
electron transport in Hall thrusters.~\cite{croes} The abnormally high electron
mobility in these simulations persists even when the electron-neutral
collisions are switched off. This enhanced transport is associated with an
electron drift instability that becomes saturated by ion trapping. After the
instability saturates, $\vec{E} \cross \vec{B}$ fluctuations result in higher
electron cross-field mobility. In addition to directly contributing to cross
field transport, one might expect significant numerical collisions to influence
both the growth rate of the instability and perhaps the magnitude of the field
fluctuations, since those are directly affected by the macroparticle weight. It
is also known that PIC can capture instability-enhanced
collisions,~\cite{scheiner} so it is necessary for the macroparticle shape to
resolve the wave numbers of the contributing instabilities. The estimate for
the numerical collision time is on the order of the saturation time of the
instability $\sim 2 \mu$s, suggesting that numerical collisions could possibly
influence the saturation, or prohibit the simulation from probing the effect of
weaker electron-neutral collision processes.

Finally, we consider two examples that emphasize how it is particularly
challenging to extend the numerical thermalization time beyond the physical
rate in 3D. The formation and evolution of plasma streamers has been studied
with 3D PIC.~\cite{teunissen} These simulations study the size and growth of
the streamers emerging from a sharp pin electrode. Using estimations for the
temperature in the streamer head (about 5 eV~\cite{taccogna2018JOPD}) the
numerical thermalization time was found to be shorter than the physical
value. It would then be expected to influence collisional transport processes,
including electrical and thermal conductivities.

The ITER fusion experiment requires a specially designed neutral beam injector
that is supplied by a high current negative ion source. The plasmas generated
in the ion source have been modeled
extensively~\cite{wunderlich,montellano,wunderlichPSST2014} including with 3D
PIC simulation~\cite{mochalskyy} due to the crossed magnetic field geometry.
The extracted negative ion flux depends strongly on the steady state plasma
screening and sheath physics near the extraction apparatus. Collision times are
important for determining the properties of the sheath and presheath, and in
these simulations, the numerical collision rate is expected to exceed the
physical rate.

\section{Conclusions}

A simple PIC code was used to show that numerical thermalization in PIC
simulations is accurately predicted by kinetic theory, particularly in relation
to the macroparticle weight, size, and dimensionality of the simulation.
Agreement with the models developed by Okuda, Birdsall,~\cite{okudaPOF1970} and
Langdon~\cite{langdonJOCP1970} based on the Lenard-Balescu collision operator
was established and the subsequent constraints on modern PIC simulations were
discussed. The kinetic theory may be used to predict the thermalization time in
PIC simulations with a variety of conditions, dimensions, and
interpolation/solver schemes. As PIC methods develop further, the force kernel
could be adapted to develop the scaling laws necessary to maintain the
physicality of simulations. Fully kinetic simulations often involve the
interaction of a variety of physical models and processes, so it is crucial to
contextualize the numerical thermalization processes for each particular
application. Some applications were discussed, in particular the various kinds
of constraints (timescale, force resolution, dispersion, \textit{etc}.) one might
encounter when specifying an informed PIC simulation. Since the shape effects
that nominally reduce collisions are mediated through the inter-particle force,
there is an inherent trade-off between the mitigation of numerical
thermalization and the accuracy of the underlying collisionless
(Vlasov) processes.

\begin{acknowledgments}
This work was supported by Sandia National Laboratories. Sandia National
Laboratories is a multi-mission laboratory managed and operated by National
Technology and Engineering Solutions of Sandia, LLC., a wholly owned subsidiary
of Honeywell International, Inc., for the U.S. Department of Energy’s National
Nuclear Security Administration under Contract No. DE-NA0003525.
\end{acknowledgments}

\section*{Author declarations}

\subsection*{Conflicts of Interest}

The authors have no conflicts to disclose.

\subsection*{Author Contributions}

\textbf{Ryan Park}:
Formal analysis (lead);
investigation (lead);
methodology (equal);
software (lead);
visualization (lead);
writing -- original draft preparation (lead);
writing -- review \& editing (equal).
\textbf{Christopher Moore}:
Conceptualization (equal); 
funding acquisition (lead);
project administration (equal);
writing -- review \& editing (equal).
\textbf{Scott Baalrud}:
Conceptualization (equal); 
formal analysis (supporting);
investigation (supporting);
methodology (equal);
project administration (equal);
supervision (lead);
writing -- review \& editing (equal).

\section*{Data Availability Statement}
The data that support the findings of this study are available from the
corresponding author upon reasonable request. 

\appendix

\section{PIC forces}
\label{app:PIC_forces}

The calculation of PIC forces can be intuitively represented using Hockney's
method of generalized functions.~\cite{hockney_gen} In this approach, full
Fourier transforms are applied over the continuous particle quantities and
all discrete quantities are represented using impulse functions. This allows
for direct connection between the algorithmic details of a given PIC
implementation and the kinetic equations discussed in Sec.~\ref{theory}.
Though the calculation assumes a uniform grid, the characteristic width of
an unstructured mesh in a particular region of simulation could be used in this
analysis to study the local thermalization behavior so long as the grid
dimensions do not vary too rapidly compared to $\lambda_\mathrm{D}$ or the support of
$S\p{\vec{r}}$.

Consider the Dirac comb (or sampling function) defined as
\begin{equation}
    \Pii\p{\vec{r}} = \sum_{\vec{g}}^{} \delta^{\p{d}}\p{\vec{r} - \vec{r}_{\vec{g}}}.
\end{equation}
$d$ is the dimensionality of the simulation, $\vec{g}$ are indices for the mesh
points, and $\vec{r}_{\vec{g}}$ are the mesh point locations with Cartesian spacing
$\Delta$. Multiplication by $\Pii$ represents the reduction of a continuous
function to a discrete set of data available only at the grid points. Next,
consider the unit top hat function with a cutoff at the grid Nyquist frequency
\begin{equation}
    \Pi\p{\vec{k}} =
    \begin{cases}
        1, & \abs{k_{\sigma}} < \pi / \Delta, \sigma \in \cu{1, ..., d}, \\
        0, & \text{otherwise}. 
    \end{cases}
\end{equation}
Multiplication by $\Pi$ in $\vec{k}$-space represents the frequency cutoff
imposed by the finite grid, i.e., the grid cannot store any information with
spatial frequency $\abs{\vec{k}} > \pi/\Delta$. With these tools, the forward
and reverse Fourier operators are discretized like so
\begin{equation}
    \mathcal{F} \to \Pi \mathcal{F} \Pii, \qquad
    \mathcal{F}^{-1} \to \Pii \mathcal{F}^{-1} \Pi.
\end{equation}
In a momentum-conserving PIC simulation, the force field calculated from a set
of particle data $\vec{r}_{j}$ with corresponding density
$n_{\delta s}\p{\vec{r}} = \lambda_{\rm D}^{3-d}W_{s} \sum_{j}^{N} \delta^{\p{d}}\p{\vec{r} - \vec{r}_{j}}$
can be written as the following operator chain
\begin{equation}
    \vec{F}_{ss'} = W_{s'} \frac{q_{s} q_{s '}}{\epsilon_{0}}
    S * \Pii \mathcal{F}^{-1} \Pi \varphi \frac{-i \vec{k}}{k^{2}} \mathcal{D} \Pi \mathcal{F} \Pii S * n_{\delta s}.
    \label{Fss_intuit}
\end{equation}
Here, $S$ is the particle shape function used both to interpolate particle data
to the grid and interpolate grid quantities to the particles. The
$\mathcal{O}\ *$ notation represents convolution of a function $\mathcal{O}$
with the entire expression right of its placement. $\varphi$ represents an
arbitrary filter, which could be implemented either in $\vec{k}$-space directly
or by discrete convolution of a grid quantity in real space. Since the field
equations are solved on the grid, they must be discretized. The common
replacement for the derivative operator is the finite difference form
\begin{equation}
    \grad \to \vec{H}\p{\Delta}*, \quad \quad \grad^{2} \to \vec{H}\p{\Delta/2} * \vec{H}\p{\Delta/2} *
    \label{grad_disc}
\end{equation}
where
\begin{equation}
    H_{i}\p{h} = \frac{\delta\p{r_{i} - h} - \delta\p{r_{i} + h}}{2h}.
\end{equation}
By inserting these discrete forms into the Poisson and Gauss field equations,
it can be seen upon Fourier transform that the electric field $\vec{E}$
and charge density $\rho$ are related by
\begin{equation}
    \vec{E}_{ss'}\p{\vec{k}} = - \frac{q_{s'}}{\epsilon_{0}}
    \frac{i\vec{k}}{k^{2}}
    \mathcal{D}\p{\vec{k}, \Delta} \rho_{s}\p{\vec{k}}
    \label{Ess}
\end{equation}
where $\mathcal{D}$ is the symmetric gradient discretization tensor often
implemented directly in FFT-based field solvers.~\cite{birdsall} In the case of
Eq.~\eqref{grad_disc},
\begin{equation}
    \mathcal{D}_{ij} = \delta_{ij}  \frac{\text{sinc}\p{k_{j} \Delta}}{\sum_l\text{sinc}^{2} \p{k_{l} \Delta/2}}.
    \label{Dtensor_App}
\end{equation}
Upon successive application of the convolution theorem,
$f~*~g~=~\mathcal{F}^{-1} \cu{\p{\mathcal{F} f} \p{\mathcal{F} g}}$, one may
commute the Fourier transforms of Eq.~\eqref{Fss_intuit} through
all other operators in pursuit of identifying $\vec{K}$ in the form of
Eq.~\eqref{Kdef},
\begin{equation}
    \vec{F}_{ss'} = W_{s '} \frac{q_{s} q_{s '}}{\epsilon_{0}} \mathcal{F}^{-1}
    S \Pii * \Pi \varphi \frac{-i \vec{k}}{k^{2}} \mathcal{D} \Pii * S \mathcal{F} n_{s '}.
\end{equation}
The operator $\Pii * \Pi$ has no bearing on the calculation, since it only
replicates the $\vec{k}$-space outside the Nyquist cutoff (which can never be
stored in the grid anyway). Using this fact and $\Pi\ \Pi = \Pi$, the PIC force
kernel is identified via~\eqref{Kdef} as
\begin{equation}
    \vec{K} = - \frac{i \vec{k}}{k^{2}} \varphi \mathcal{D}S \Pii * S.
    \label{eq:reduced_pic_kernel}
\end{equation}
For details regarding
energy-conserving PIC, see Ref.~\onlinecite{langdonJOCP1973} for the Lagrangian derivation of
the original energy-conserving algorithm and Ref.~\onlinecite{lewis} for a
discussion and derivation of the EC-PIC force kernel that is relevant in place
of Eq.~\eqref{eq:reduced_pic_kernel},
\begin{equation}
    \vec{K} = - \frac{i \vec{k} \varphi S \Pii * S}{\p{\Pii * k^{2} S^{2}}}.
\end{equation}

\section{1D thermalization}
\label{app:1D_thermalization}

The VDF of each individual species will evolve toward the
total mass weighted distribution in 1D simulations~\cite{eldridge}
\begin{equation}
    \lim_{N_{\text{D}} < t \omega_{p} < N_\mathrm{D}^{2}} \br{m_{s} f_{s}} \propto \sum_{s'}^{} m_{s'} f_{s'}
\end{equation}
on timescales $t \omega_{p} \sim N_{\text{D}}$ and towards a Maxwellian on
timescales $t \omega_{p} \sim N_{\text{D}}^{2}$. Though it may be dubious to
regard the first-order process as ``thermalization'', the constraints for
multispecies simulations are the same. Simulations are expected to accurately
describe non-Maxwellian behavior and so should not become artificially
Maxwellian. Equivalently, simulations are expected to model the VDF of each
participating species and so each distribution should not become artificially
proportional to $\sum_{s '}^{} m_{s '} f_{s '}$. For example, the simulated
drag on an electron beam streaming against a Maxwellian background of ions in
1D would be of the same order as an equivalent beam in 2D
or 3D since $m_\mathrm{ion} f_{\text{ion}} + m_\mathrm{e} f_\mathrm{e} \approx
m_{\text{ion}} f_\mathrm{ion}$ and thus the respective drift and diffusion
coefficients $\vec{A}\p{\vec{v}}$, $\mathbb{D}\p{\vec{v}}$ and their associated
relaxation timescales are still the relevant metrics for VDF evolution via
collisions.
The $1/N_\mathrm{D}^{2}$ scaling property of 1D simulations is also notably violated
when Monte Carlo collisions are included, as discussed by
Turner.~\cite{turner,jubinPOP2024}

\bibliography{refs}	

\end{document}